\documentclass[useAMS,usenatbib]{mnras}

\usepackage{mathtools}

\usepackage{graphicx,epsfig,times,color,amsmath,amssymb,pifont}

\usepackage{subfig}

\usepackage{amssymb} 

\usepackage{url}

\usepackage{aas_macros}

\def\Msun{\mbox{~M$_\odot$}}

\def\Msunns{\mbox{M$_\odot$}}
\def\Msunpc2{\mbox{~M$_\odot$~pc$^{-2}$}}
\def\Msunyr{\mbox{~M$_\odot$~yr$^{-1}$}}
\def\kms{\mbox{~km~s$^{-1}$}}

\def\pc{\mbox{~pc}}
\def\kpc{\mbox{~kpc}}
\def\Mpc{\mbox{~Mpc}}

\def\Myr{\mbox{~Myr}}
\def\erg{\mbox{~erg}}
\def\s{\mbox{~s}}

\def\tff{t_{\rm ff}}
\def\tfb{t_{\rm SF}}
\def\tsn{t_{\rm OB}}
\def\tsnO{t_{\rm OB,0}}

\def\vdispISM{\sigma_{\rm ISM}}

\def\rhoc{\rho_{\rm c}}
\def\phifb{\phi_{\rm fb}}
\def\phiP{\phi_{\rm P}}
\def\Mcloud{M_{\rm cloud}}
\def\Mc{M_{\rm c}}
\def\epsbound{\epsilon_{\rm bound}}
\def\epscore{\epsilon_{\rm core}}
\def\epsfb{\epsilon_{\rm SF}}
\def\epsff{\epsilon_{\rm ff}}

\def\fSigma{f_{\Sigma}}
\def\SigmaISM{\Sigma_{\rm ISM}}
\def\Sigmac{\Sigma_{\rm c}}
\def\MOB{M_{\rm OB}}
\def\Mth{M_{\rm th}}
\def\Mmin{M_{\rm min}}
\def\Mmax{M_{\rm max}}
\def\Mc{M_{\rm c}}

\def\Rc{R_{\rm c}}
\def\Gyr{~\rm Gyr}

\def\apj{\rm Astrophys. J.}
\def\aap{\rm Astron. \&Astrophys. }
\def\mnras{\rm Mon.\,Not.\,RAS. }

\def\yr{{\rm yr}\ }
\def\mathnew{\mathsurround=0pt}
\def\simov#1#2{\lower .5pt\vbox{\baselineskip0pt
    \lineskip-.5pt\ialign{$\mathnew#1\hfil##\hfil$\crcr#2\crcr\sim\crcr}}}

\def\apj{\rm ApJ}
\def\apjl{\rm ApJL}
\def\apjs{\rm ApJS}
\def\aj{\rm AJ}
\def\mnras{\rm MNRAS}
\def\nat{\rm Nature}

\def\aap{\rm A\&A}
\def\araa{\rm ARAA}
\def\pasp{\rm PASP}

\newcommand{\HI}{\hbox{{\sc H}\hspace{0.7pt}{\sc i}} }

\defcitealias{Reina-Campos17}{RK17}

\title[The minimum mass of stellar clusters]{A model for the minimum mass of bound stellar clusters and its dependence on the galactic environment}
\author[S.~Trujillo-Gomez, M.~Reina-Campos \& J.~M.~D.~Kruijssen]
{Sebastian Trujillo-Gomez\thanks{E-mail: strujill@gmail.com}, 
Marta Reina-Campos
and J.~M.~Diederik Kruijssen \\ 
Astronomisches Rechen-Institut, Zentrum f{\"u}r Astronomie der Universit{\"a}t Heidelberg, Monchhofstra{\ss}e 12-14, D-69120 Heidelberg, Germany
}

\begin{document}

\date{Accepted 2019 July 9. Received 2019 June 17; in original form 2019 March 11}

\pagerange{\pageref{firstpage}--\pageref{lastpage}} \pubyear{2019}

\maketitle

\label{firstpage}

\begin{abstract}
We present a simple physical model for the minimum mass of bound stellar clusters as a function of the galactic environment. The model evaluates which parts of a hierarchically-clustered star-forming region remain bound given the time-scales for gravitational collapse, star formation, and stellar feedback. We predict the initial cluster mass functions (ICMFs) for a variety of galaxies and we show that these predictions are consistent with observations of the solar neighbourhood and nearby galaxies, including the Large Magellanic Cloud and M31. In these galaxies, the low minimum cluster mass of $\sim10^2\Msun$ is caused by sampling statistics, representing the lowest mass at which massive (feedback-generating) stars are expected to form. At the high gas density and shear found in the Milky Way's Central Molecular Zone and the nucleus of M82, the model predicts that a mass $>10^2\Msun$ must collapse into a single cluster prior to feedback-driven dispersal, resulting in narrow ICMFs with elevated characteristic masses. We find that the minimum cluster mass is a sensitive probe of star formation physics due to its steep dependence on the star formation efficiency per free-fall time. Finally, we provide predictions for globular cluster (GC) populations, finding a narrow ICMF for dwarf galaxy progenitors at high redshift, which can explain the high specific frequency of GCs at low metallicities observed in Local Group dwarfs like Fornax and WLM. The predicted ICMFs in high-redshift galaxies constitute a critical test of the model, ideally-suited for the upcoming generation of telescopes.
\end{abstract}


\section{Introduction}
\label{sec:intro}

Star clusters are potentially powerful tools for understanding the assembly of galaxies in a cosmological context. For instance, the properties of globular cluster (GC) systems are tightly correlated not only to their host galaxies \citep{BrodieStrader06,Kruijssen14}, but also to their inferred dark matter halo masses \citep{Blakeslee97,Harris17,Hudson18}. Likewise, young stellar clusters provide detailed information about the recent star formation conditions in their host galaxies \citep{PortegiesZwart10,Longmore14,Chilingarian18}. Clusters are also ideal tracers of gravity and probe the detailed mass distribution of dark matter haloes \citep{Cole12,Erkal15,Alabi16,Contenta18,vanDokkum18b}. In order to fully exploit stellar clusters as tracers of galaxy and structure formation, we must understand how their birth environments give rise to their initial properties (including masses, ages, structure and chemical composition) and how these evolve across cosmic time. 

Understanding the relation between star and cluster formation has important implications for the hierarchical formation and evolution of galaxies. A common hypothesis for the origin of GCs considers them products of regular star formation in the extreme conditions in the interstellar medium (ISM) of $z \sim 2-3$ galaxies \citep[e.g.][]{KravtsovGnedin05, Elmegreen10, Shapiro10, Kruijssen15b, Reina-Campos19}. Within this framework, GCs correspond to the dynamically-evolved remnants of massive clusters formed at high redshift \citep[e.g.][]{Forbes18,Kruijssen19a}. Recently studies are showing that GCs are excellent tracers of the assembly histories of galaxies, and in particular of the Milky Way \citep[e.g.][]{Kruijssen18b,Myeong18}. Naturally, the properties of GC populations can only be predicted given a complete model for their initial demographics, of which the initial cluster mass function (ICMF) is an essential component.

Many fundamental aspects of the process of star cluster formation are still poorly understood, including the fraction of stars that form in clusters, and the ICMF, and their dependence on the large-scale galactic environment. For example, it is a widespread assumption in the literature that the initial mass function of bound star clusters follows a power-law with a logarithmic slope $\sim -2$ \citep{ZhangFall99, Bik03, Hunter03, McCradyGraham07, Chandar10, PortegiesZwart10}. This was later revised to include a high-mass truncation \citep[cf.][]{Schechter76}, with additional evidence of a strong environmental dependence of the truncation mass \citep[e.g.][]{Gieles06, Larsen09, Adamo15, Johnson17, Reina-Campos17, Messa18}. Despite all the effort that has been put into understanding and modelling the environmental dependence of the high-mass end, the low-mass truncation is still assumed to be $\sim 10^2\Msun$ \citep[e.g.][]{LadaLada03, Lamers05} across all environments. 

Recently, \citet[][hereafter \citetalias{Reina-Campos17}]{Reina-Campos17} developed a model for the maximum mass of stellar clusters that simultaneously includes the effect of stellar feedback and centrifugal forces. The model predicts the upper mass scale of molecular clouds (and by extension, that of star clusters) by considering how much mass from a centrifugally-limited region (containing a `Toomre mass', see \citealt{Toomre64}) can collapse before stellar feedback halts star formation. The authors find that the resulting upper truncation mass of the ICMF depends on the gas pressure, where environments with higher gas pressures are able to form more massive clusters. Local star-forming discs (such as the Milky Way and M31) with low ISM surface densities are predicted to have much lower truncation masses than (nuclear) starbursts and $z \sim 2$ clumpy discs, where stellar feedback is slow relative to the collapse time-scale. Together with a theoretical model predicting an increase of the cluster formation efficiency (CFE) with gas pressure \citep{Kruijssen12b}, these results reproduce observations of young massive clusters (YMCs) in the local Universe \citep[e.g.][]{Adamo15,Messa18}. The general implication of these results is that cluster properties are shaped by the galactic environment. This environmental coupling hints at the exciting prospect of using clusters to trace the evolution of their host galaxies.  

In addition to allowing the use of clusters as tracers of galaxy assembly, the above models also provide the initial conditions for studies of cluster dynamical evolution \citep[e.g.][]{LamersGieles06, Lamers10, Baumgardt19}, as well as for sub-grid modelling of star cluster populations in cosmological simulations. Together with the environmentally-dependent modelling of dynamical evolution including tidal shocking and evaporation, these cluster formation models enable the formation and evolution of the entire star cluster population to be followed from extremely high redshift down to $z=0$ \citep{Pfeffer18, Kruijssen19a}. Self-consistently forming and evolving the entire cluster population in cosmological simulations of a representative galaxy sample is currently an intractable problem due to the extremely high resolution required, although case studies are promising \citep{kim18,li18}.

Despite the recent progress on the theory of GC formation, many of the observed properties of GC populations still remain a puzzle. The GC mass function has a close to log-normal shape with a characteristic peak at $\sim 10^5\Msun$ \citep[e.g.][]{Harris91, Jordan07}, whereas the young cluster mass function (CMF) continues as a power law down to much lower masses \citep[e.g.][]{ZhangFall99, Hunter03, Johnson17}. This difference can be explained if the majority of low-mass clusters are disrupted over several Gyr due to dynamical effects \citep{Spitzer87,Gnedin99,FallZhang01,BaumgardtMakino03,LamersGieles06,Kruijssen15b}. However, recent observations of GC systems in nearby dwarf galaxies seem to challenge this scenario.

\citet{Larsen12, Larsen14} determined the chemical properties of GCs around a number of Local Group dwarf galaxies. These studies found that a strikingly large fraction ($\sim 20-50$~per cent) of low metallicity stars in Fornax and in WLM belong to their GCs (which have a characteristic mass of $\sim 10^5\Msun$). This is extremely high compared to the typical fraction of $0.1$~per cent found in Milky Way-mass galaxies, and it is also the largest GC specific frequency ever observed. This feature seems to extend to every dwarf galaxy where GC and field star metallicities have been determined, and contradicts the existence of a universal power-law ICMF down to a common lower mass limit of $\sim10^2\Msun$ \citep{Larsen18}. In these dwarf galaxies, the traditionally assumed universal \citep{Schechter76} ICMF extending down to $\sim 10^2\Msun$ requires the majority of the low-mass clusters to have been disrupted after a Hubble time of dynamical evolution, thus returning their mass to the field population. This would allow at most $10$~per cent of the low-metallicity stars to reside in the surviving GCs, contrary to the much larger observed fraction of $20{-}50$~per cent. 

In this paper, we examine the possibility that the low-mass end of the ICMF is not universal, but is instead determined by the environmentally dependent minimum mass of a bound star cluster. We develop a model for the dependence of the minimum cluster mass on galactic birth environment. The model is based on the hierarchical nature of star formation in molecular clouds regulated by stellar feedback, combined with empirical input on the structure and scaling relations of clouds in the local Universe. By estimating the time-scale for stellar feedback to halt star formation in relation to the collapse time of clouds with a spectrum of masses, we can predict the range of cloud masses that can achieve the minimum star formation efficiency needed to remain bound after the remaining gas is blown out by feedback. This minimum mass scale emerges naturally as the largest scale that must collapse and merge into a single bound object, which corresponds to the bottom of the hierarchy of young stellar structure in galaxies.

The paper is organised as follows. In Section~\ref{sec:model}, we present the derivation of the minimum bound cluster mass as a function of cloud properties as well as global galaxy observables. Section~\ref{sec:uncertainties} presents an estimate of the dominant uncertainties. Section~\ref{sec:environments} illustrates the predicted variation of the minimum mass and the width of the ICMF across the broad range of observed galaxies. In Section~\ref{sec:predictions}, we make predictions of the full ICMF and compare these with observational estimates in the solar neighbourhood, the Large Magellanic Cloud (LMC), M31, the Antennae galaxies, and galactic nuclei including the Central Molecular Zone (CMZ) of the Milky Way, and the nucleus of M82. This section also discusses the effect of the minimum mass on the inferred CFE. In Section~\ref{sec:GCs}, we illustrate how the model can be used to reconstruct the galactic environment that gave rise to the populations of GCs in the Fornax dSph, and also to predict the ICMFs in the high-redshift environments that will be within reach of the next generation of observational facilities. Lastly, Section~\ref{sec:conclusions} summarises our results.

\section{Model}
\label{sec:model}

We begin by assuming that the ICMF follows a power law with exponential truncations at both the high- and low-mass ends:
\begin{equation}
    \frac{{\rm d}N}{{\rm d}M} \propto M^{\beta} \exp\left( -\frac{\Mmin}{M} \right) \exp\left( -\frac{M}{\Mmax} \right) ,
    \label{eq:CMF}
\end{equation}
where $\beta = -2$ as expected from gravitational collapse in hierarchically structured clouds \citep[e.g.][]{Elmegreen96,Guszejnov18}, $\Mmin$ is the minimum cluster mass scale, and $\Mmax$ is the maximum cluster mass scale, which we determine from the mass of the largest molecular cloud that can survive disruption by feedback or galactic centrifugal forces, according to the model by \citetalias{Reina-Campos17}. This ICMF introduces three different regimes. At $M\gg\Mmax$, bound clusters are extremely unlikely to form due to the disruptive effects of galactic dynamics and stellar feedback, inhibiting the collapse of the largest spatial scales. At $M\ll\Mmin$, bound clusters must be part of a larger bound part of the hierarchy, because the attained star formation efficiencies are very high, causing them to merge into a single bound object of a higher mass. In between these mass scales, self-similar hierarchical growth imposes a power law ICMF.

When describing star and cluster formation in a disc in hydrostatic equilibrium \citep[cf.][]{krumholzmckee05,Kruijssen12b}, $\Mmax$ can be expressed in terms of the ISM surface density $\SigmaISM$, the angular velocity of the rotation curve $\Omega$, and the Toomre $Q$ parameter of the galactic gas disc. In this section, we outline a model to derive the minimum bound cluster mass $\Mmin$ and in Section~\ref{sec:minmass} we present the analytical formalism. In Section~\ref{sec:IMF}, we include the impact of sampling the stellar initial mass function (IMF) in low-mass molecular clouds, and in Section~\ref{sec:global} we formulate the minimum mass in terms of global galaxy properties.

Following \citet{Kruijssen12b} and \citetalias{Reina-Campos17}, we model star and cluster formation as a continuous process that takes place when overdense regions within molecular clouds and their substructures collapse due to local gravitational instability. The collapse leads to fragmentation and the formation of stars until the newly formed stellar population deposits enough feedback energy and momentum in the local gas reservoir to stop the gas supply and the corresponding star formation.

The critical time-scale that defines how much gas is converted into stars is determined by the time required for stellar feedback to halt star formation. This `feedback time-scale' determines the total star formation efficiency through the relation
\begin{equation}
    \epsfb \equiv \frac{M_*(t_{\rm SF})}{\Mc} ,
    \label{eq:epsfb1}
\end{equation}
where $M_*(t_{\rm SF})$ is the mass of stars formed after a feedback time-scale $t_{\rm SF}$ and $\Mc$ is the cloud mass. This feedback-regulated star formation efficiency also determines the ability of the star cluster to stay bound after the residual gas is expelled. The detailed role of various stellar feedback processes in cloud disruption is still a highly debated topic in the literature \citep{Korpi99, JoungMacLow06, Thompson05, Murray10, Dobbs11, Dale12, Kruijssen19b}. These processes include photoionisation, radiation pressure, stellar winds, and supernova (SN) explosions. The detailed treatment of each of these processes is well beyond the scope of our model. We therefore use the fact that the integrated specific momentum output of each of these mechanisms is similar \citep[e.g.][]{Agertz13} and use feedback by SN explosions as a phenomenological proxy for the complete array of feedback processes. In what follows, we will use the the minimum lifetime of O- and B-type stars as the time delay until the first SNe, $t_{\rm OB} \simeq 3\Myr$. This is strictly an upper limit on the delay for the onset of the energetic effects of feedback. We refer the reader to Section~\ref{sec:earlyfeedback} for a discussion of the uncertainties related to this assumption.

To obtain the minimum star cluster mass within a galaxy with a given set of characteristic global properties (such as the ISM surface density and the angular velocity), we must calculate the range of cloud masses in which the star formation efficiency is guaranteed to be large enough for the stars to collapse into a single cluster that remains bound after the remaining gas is expelled by stellar feedback \citep{Hills80, Lada84, Kroupaetal01}. Motivated by the comprehensive exploration of parameter space in idealised $N$-body simulations \citep[e.g.][]{BaumgardtKroupa07}, this condition can be written in terms of the minimum local star formation efficiency needed to form a bound cluster ($\epsilon_{\rm min}$) as
\begin{equation}
    \epsfb \geq \epsilon_{\rm min} ,
    \label{eq:epsfb2a}
\end{equation}
such that star-forming regions that do not convert at least half of their gas mass into stars by the time gas is expelled will not form bound clusters. Because the maximum star formation efficiency is limited by feedback from protostellar outflows disrupting protostellar cores (the formation sites of individual stars), which is adiabatic and therefore undisruptive, this condition can be written
\begin{equation}
    \epsfb \geq \epsbound\epscore ,
    \label{eq:epsfb2}
\end{equation}
where $\epscore$ is the limiting efficiency of star formation within protostellar cores, and $\epsbound$ is the minimum fraction of cloud mass that must condense into molecular cores to obtain a bound cluster. This expression is one of the key ingredients of our model. Idealised $N$-body simulations find values $\epsbound \approx 0.4$ across a broad variety of cluster properties and environments \citep{BaumgardtKroupa07}. Observations of protostellar cores find $\epscore \approx 0.5$ \citep[e.g.][]{Enoch08}, such that the star formation efficiency should be $\epsfb \geq 0.2$ in order to guarantee collapse into a single bound cluster.

Summarising, the procedure to obtain the minimum bound cluster mass is as follows.
\begin{enumerate}
    \item Derive the total star formation time-scale, $\tfb$ from the time required for stellar feedback to over-pressure a collapsing gas cloud of density $\rhoc$ embedded in a galactic disc with ISM surface density $\SigmaISM$.
    \item Express the mean volume density in terms of the cloud surface density and mass by assuming spherical symmetry.
    \item Obtain the total star formation efficiency of the cloud, $\epsfb$, as a function of cloud mass and surface density by multiplying the ratio of the star formation time-scale and the free-fall time by the empirical star formation efficiency per free-fall time. Because low mass clouds have higher gas densities, the integrated star formation efficiency will be a decreasing function of cloud mass.
    \item Compare the total star formation efficiency ($\epsfb$) to the efficiency required for the cluster to remain bound after stellar feedback blows out the residual gas ($\epsbound\epscore$). The maximum cloud mass that reaches this threshold efficiency will set the minimum scale for collapse into a single bound cluster, because {\it lower mass scales are part of a larger bound part of the hierarchy}, i.e.~they are guaranteed to merge into larger bound structures before star formation is halted by feedback. This scale then defines the bottom of the merger hierarchy.
    \item The minimum bound cluster mass as a function of cloud mass, cloud surface density, and ISM surface density is then obtained by multiplying the threshold cloud mass by the minimum required efficiency to remain bound, $\epsbound\epscore$.
\end{enumerate}
In the following section, we follow the above procedure to derive the minimum cluster mass.

\subsection{The minimum mass of a bound star cluster}
\label{sec:minmass}

Finding the minimum star cluster mass amounts to solving equations (\ref{eq:epsfb1}) and (\ref{eq:epsfb2}) simultaneously for $\Mmin = M_*$. In other words, we must find the range of cloud (and resulting cluster) masses where star formation is efficient enough to reach $\epsfb \geq \epsbound\epscore$, such that the local stellar population is guaranteed to remain gravitationally bound, even after any residual gas reservoir is expelled by stellar feedback. Because the star formation efficiencies are defined locally, we are interested in the largest mass scale at which boundedness is certain to be achieved. Lower-mass aggregates will be part of a larger bound structure, such that the minimum cluster mass is set by the largest structure that {\it must} be gravitationally bound.

The first step is to obtain the feedback-regulated star formation efficiency as a function of the cloud mass. The integrated local star formation efficiency can be expressed in terms of the specific star formation efficiency per free-fall time as \citep{Kruijssen12b}
\begin{equation}
    \epsilon_{\rm SF} = \epsff \frac{ t_{\rm SF} }{ t_{\rm ff} } ,
    \label{eq:5}
\end{equation}
where $t_{\rm SF}$ is the total duration of the star formation process in the cloud, and $\epsff$ and $\tff$ are the star formation efficiency per free-fall time and the mean cloud free-fall time, respectively. Motivated by detailed measurements of molecular clouds in the Milky Way \citep[e.g.][]{Evans14,Lee16}, as well as  across many nearby galaxies \citep{Leroy17,Utomo18}, we assume a fiducial constant value $\epsff = 0.01$ (see Section~\ref{sec:uncertainties} for a discussion of the uncertainty on this number).

Following \citet{Kruijssen12b}, the duration of star formation within a gas reservoir of density $\rhoc$, i.e.\ $\tfb$, is set by the time it takes for stellar feedback to pressurise the gas and stop the supply of fresh material. This time-scale can be calculated by comparing the external confining pressure of the ISM (or parent molecular cloud) to the gas pressure within the feedback-affected region. The duration of star formation is then obtained by adding the time delay between the onset of star formation ($\tsn$) and the first SN explosion and the time between the first SN and pressure equilibrium with the ISM ($t_{\rm eq}$), i.e.
\begin{equation}
    \label{eq:t_fb_def}
    \tfb = \tsn + t_{\rm eq} .
\end{equation}
We start by writing the the ambient pressure of the ISM at the disc midplane as
\begin{equation}
    \label{eq:pressure1}
    P_{\rm ISM} = \phi_{\rm P} \frac{\pi}{2} G ~\SigmaISM^2 ,
\end{equation}
where $\SigmaISM$ is the surface density of the ISM of the galaxy, and $\phi_{\rm P} \approx 3$ is a correction due to the gravity of the stars \citep{krumholzmckee05}. The outward pressure exerted by stellar feedback is \citep{Kruijssen12b}
\begin{align}
  \label{eq:pressure2}
    P_{\rm fb} & = \frac{ E_{\rm fb} }{ V } \nonumber \\
               & = \phifb \epsfb \rhoc ~t_{\rm eq} ,
\end{align}
where $E_{\rm fb}$ is the feedback energy per unit stellar mass, $V$ is the region volume, $\rhoc$ is the mean cloud density, $\phi_{\rm fb} = 3.2 \times 10^{32} \erg \s^{-1} \Msun^{-1}$ is the mean rate of SN energy injection per unit stellar population mass , $\epsfb$ is the total fraction of gas mass that is converted into stars during the lifetime of the cloud (equation~\ref{eq:5}), and $t_{\rm eq}$ is the time it takes for the cloud to reach pressure equilibrium with the ISM after the death of the first massive (OB-type) star. 

By equating the inward and outwards pressures we can then solve for the time required to reach pressure equilibrium and stop gas accretion $t_{\rm eq}$ using equations (\ref{eq:pressure1}) and (\ref{eq:pressure2}),
\begin{equation}
    \label{eq:t_eq}
    t_{\rm eq} = \frac{ \pi \phi_{\rm P} G ~\SigmaISM^2 }{ 2 \phifb \epsfb \rhoc } .
\end{equation}
The time-scale corresponding to the duration of star formation, $\tfb$, may then be defined as the time required for stellar feedback to cut the fresh gas supply and halt star formation. Using equations~(\ref{eq:t_fb_def}), (\ref{eq:pressure1}), and (\ref{eq:pressure2}), we obtain
\begin{equation}
    \label{eq:2}
    \tfb = \tsn + \frac{ \pi \phiP G ~\SigmaISM^2 }{ 2 \phifb \epsfb \rhoc } ,
\end{equation}
where $\tsn$ is the time delay before the first SN (corresponding to an OB-type progenitor star) explodes. 

The next step is to express the molecular cloud density in terms of its mass and mean surface density. Assuming spherical symmetry, the mean cloud gas density is
\begin{equation}
    \rhoc = \frac{ \Mc }{ \frac{4}{3}\pi\Rc^3 }  .
    \label{eq:rho_c_rad}
\end{equation}
Expressing the cloud radius in terms of the mean cloud surface density, $\Sigmac$, yields
\begin{equation}
    \label{eq:mean_sd}
    \Rc = \left( \frac{ \Mc } { \pi \Sigmac } \right)^{1/2}.
\end{equation}
Substituting this expression into equation (\ref{eq:rho_c_rad}) gives
\begin{equation}
    \rhoc = \frac{3}{4} \left( \frac{ \pi \Sigmac^3 }{ \Mc } \right)^{1/2} .
    \label{eq:rho_c}
\end{equation}

The final expression for the duration of star formation $\tfb$ can now be obtained in terms of the cloud mass and surface density by substituting equations (\ref{eq:5}) and (\ref{eq:rho_c}) into equation~(\ref{eq:2}),
\begin{equation}
    \tfb = \tsn + \frac{2\pi}{3} \frac{ \phiP G ~\SigmaISM^2 \tff }{ \phifb \epsff \tfb} \left(\frac{ \Mc }{ \pi \Sigmac^3 }\right)^{1/2},
    \label{eq:t_sf_quad}
\end{equation}
where the free-fall time can be written in terms of the cloud mass and surface density using equation~(\ref{eq:rho_c}), i.e.
\begin{equation}
    \tff = \sqrtsign{ \frac{3 \pi}{32 G \rhoc} } =  \sqrtsign{ \frac{\pi^{1/2}}{8G} } \left( \frac{\Mc}{\Sigmac^3} \right)^{1/4} \equiv \mathcal{C}\left( \frac{\Mc}{\Sigmac^3} \right)^{1/4} ,
    \label{eq:t_ff}
\end{equation}
where the final equality defines the constant $\mathcal{C}$.

The star-formation time-scale can now be obtained by solving the quadratic equation (\ref{eq:t_sf_quad}) after substituting equation~(\ref{eq:t_ff}) for $\tff$. The result is
\begin{equation}
    \tfb = \frac{\tsn}{2} \left[ 1 + \sqrtsign{ 1 + \frac{ 8\pi^{1/2}\mathcal{C} }{3}  \frac{ \phiP G ~\SigmaISM^2 }{ \phifb \epsff \tsn^2 }  \frac{ \Mc^{3/4} }{ \Sigmac^{9/4} }   } \right] .
    \label{eq:t_fb}
\end{equation}

At a constant cloud surface density, which is typically observed within a given galactic environment \citep[e.g.][]{Heyer09,Sun18}, two cloud mass regimes emerge. For large cloud masses, the second term inside the square root in equation~(\ref{eq:t_fb}) becomes $\gg 1$, and the star formation time-scale $\tfb$ is proportional to $\Mc^{3/4}$. Physically, this describes the regime where the time required to build up enough SN energy to pressurise the cloud increases with cloud mass, because more massive clouds have lower volume densities (and hence lower integrated star formation efficiencies and SN energy per unit cloud mass). For lower cloud masses, the second term inside the square root becomes $\ll 1$, and the feedback time-scale $\tfb \to \tsn$. This corresponds to the physical regime where the cloud mass is low enough (and hence its density and integrated star formation efficiency high enough) that the first SN provides enough energy density to overpressure the cloud and halt star formation. In this regime, the duration of star formation is then set by the time delay until the first SN by massive star formation and stellar evolution, combined with the sampling of the IMF. We will include this effect in the following section.  

Finally, we are now able to write the condition to form a bound star cluster in terms of the molecular cloud mass $\Mc$ and surface density $\Sigmac$ by substituting equations (\ref{eq:5}), (\ref{eq:t_ff}), and (\ref{eq:t_fb}) into equation~(\ref{eq:epsfb2}),
\begin{equation}
 \begin{multlined}
    \epsff \frac{\tsn}{2\mathcal{C}} 
    \frac{\Sigmac^{3/4}}{\Mc^{1/4}} 
    \left[ 1 + \sqrtsign{ 1 + \frac{ 8\pi^{1/2}\mathcal{C} }{3}  \frac{ \phiP G ~\SigmaISM^2 }{ \phifb \epsff \tsn^2 }  \frac{ \Mc^{3/4} }{ \Sigmac^{9/4} }   } \right]  \\ 
    \geq \epsbound \epscore .
 \end{multlined}
 \label{eq:boundcond}
\end{equation}

\subsection{Impact of IMF sampling on the feedback time-scale}
\label{sec:IMF}

In deriving the feedback time-scale, we implicitly excluded the effects of stochastically sampling the stellar IMF. Naturally, the derivation of the feedback time-scale, $\tfb$, in Section~\ref{sec:minmass} breaks down for arbitrarily low cloud masses, as $M \downarrow 0$ and the cloud mass becomes too small to produce even a single massive star. This should lead to a rapid rise in the delay time $\tsn$ as the cloud mass decreases, because it takes longer to build up enough cluster mass ($\MOB$) to produce a massive star. We can account for this effect by writing the characteristic feedback time-scale as
\begin{equation}
    \tsn = \tsnO + \Delta t,
    \label{eq:tsn_IMF}
\end{equation}
where $\Delta t$ is the delay from the onset of star formation until the cluster has enough mass to contain at least one massive OB-type star, and $\tsnO$ is the time interval between this moment and the onset of stellar feedback. We assume $\tsnO = 3\Myr$ based on the shortest lifetime of a star more massive than $8\Msun$ \citep[e.g.][]{Ekstrom12}.

To calculate the delay $\Delta t$ in this low-mass regime we write the minimum stellar mass $\MOB$ that must be formed for the cluster to contain at least one massive star as
\begin{equation}
    \MOB = \epsilon \Mc = \epsff \frac{\Delta t}{\tff} \Mc ,
    \label{eq:M_OB}
\end{equation}
where $\epsilon$ is the integrated star formation efficiency of the cloud. For low enough masses the mass of the cloud becomes smaller than $\MOB$ and $\Delta t > \tff/\epsff$. This corresponds to the regime of such low cloud masses that no massive stars can be produced and star formation cannot be stopped by stellar feedback.We assume that such low-mass clouds will continue to accrete until they reach sufficiently high masses to form a massive star. The value of $\MOB$ can then be obtained by solving the system of integral equations (for a given choice of stellar IMF)
\begin{align}
    \int^{\infty}_{8M_{\odot}} \Phi \frac{{\rm d}N(m)}{{\rm d}m} {\rm d}m    & = 1 ,  \\
    \int^{\infty}_{0.08M_{\odot}} \Phi \frac{{\rm d}N(m)}{{\rm d}m} m ~{\rm d}m & = \MOB,
\end{align}
for $\MOB$ and the normalisation of the IMF, $\Phi$. These two equations simply state that there is one massive star in the cluster and that the total mass under the IMF is $\MOB$. The relevant integration limits are the hydrogen-burning mass limit, $0.08\Msun$, and  the minimum mass of a B star, $8\Msun$. Solving the equations above for a \citet{chabrier03} IMF gives $\MOB = 99\Msun$. 

The time $\Delta t$ required to form at least one massive star is then obtained by inverting equation~(\ref{eq:M_OB}),
\begin{equation}
    \Delta t = \frac{\MOB}{\epsff} \frac{\tff}{\Mc} .
    \label{eq:delta_t}
\end{equation}
Substituting equation~(\ref{eq:t_ff}) to express the free-fall time, the general expression for the time delay between the onset of star formation and the first SN in a cloud of mass $\Mc$ and surface density $\Sigmac$ is
\begin{equation}
    \label{eq:t_sn}
    \tsn = \tsnO + \mathcal{C} \frac{\MOB}{\epsff} \left(\Sigmac \Mc \right)^{-3/4} .
\end{equation}
This means that the delay time increases as the cloud mass and surface density decrease, as expected from the corresponding changes of the star formation rates and free-fall times.

\subsection{Dependence on global galactic environment}
\label{sec:global}

To relate the minimum bound cluster mass to its galactic star-forming environment, we should express the condition to form a bound cluster (equation~\ref{eq:boundcond}) in terms of the properties of the host galaxy. This condition already has a dependence on the mean ISM surface density $\SigmaISM$, in addition to the dependence on the cloud mass $\Mc$ and surface density $\Sigmac$. 

The mean cloud surface density can be written as a function of the global ISM surface density in the host galaxy using equation (9) from \citet{Kruijssen15b},  
\begin{equation}
  \label{eq:f_sigma}
    f_{\Sigma} = \frac{\Sigmac}{\SigmaISM} = 3.92 \left( \frac{10 - 8f_{\rm mol}}{2} \right)^{1/2} ,
\end{equation}
where the global molecular gas fraction, $f_{\rm mol}$, is a function of the ISM surface density, $\SigmaISM$, parameterised using equation (73) of  \citet{krumholzmckee05} as
\begin{equation}
    f_{\rm mol} \approx \left[1 + 0.025~\left(\frac{\SigmaISM}{10^2\Msunpc2}\right)^{-2} \right]^{-1} .
\end{equation}
This relation implies that in galaxies with high ISM surface densities ($\SigmaISM \ga 100\Msunpc2$), the ISM becomes nearly fully molecular $f_{\rm mol} \sim 1$, and $\fSigma \sim 4$. For the low ISM surface densities characteristic of nearby spirals, $\SigmaISM \sim 10\Msunpc2$, the density contrast is $\fSigma \approx 7.7$.

The expressions for the free-fall (equation~\ref{eq:t_ff}) and the OB star (equation~\ref{eq:t_sn}) time-scales now become
\begin{equation}
  \label{eq:tff_sigma}
    \tff(\SigmaISM,\Mc) = 
    \mathcal{C} \frac{ \Mc^{1/4} }{ \left(f_{\Sigma}\SigmaISM \right)^{3/4} }  ,
\end{equation}
and 
\begin{equation}
  \label{eq:tob_sigma}
    \tsn(\SigmaISM,\Mc) = \tsnO + 
    \mathcal{C} \frac{\MOB}{\epsff}\left(f_{\Sigma} \SigmaISM \Mc \right)^{-3/4} .
\end{equation}
We may now rewrite the condition to form a bound cluster (equation~\ref{eq:boundcond}) solely in terms of the galactic ISM surface density. 

Using equations (\ref{eq:f_sigma}), and (\ref{eq:tob_sigma}),  equation~(\ref{eq:boundcond}) can be rewritten to obtain the condition for the threshold cloud mass, $\Mth$, at which the total feedback-regulated star formation efficiency, $\epsfb$, is large enough for the stars to remain bound after gas expulsion. Writing the functional dependence of the feedback time-scale implicitly we obtain
\begin{align}
    & \epsbound \epscore = \frac{\epsff}{2\mathcal{C}} \tsn(\SigmaISM,\Mth) \frac{ \fSigma^{3/4} \SigmaISM^{3/4} }{ \Mth^{1/4} }  \times \nonumber \\
    & \left[ 1 + \sqrtsign{ 1 + \frac{ 8\pi^{1/2} \mathcal{C} G \phiP }{  3 ~\phifb \epsff } \frac{ \Mth^{3/4} }{ \tsn^2(\SigmaISM,\Mth) \fSigma^{9/4} \SigmaISM^{1/4} } } \right] ,
    \label{eq:boundcondISM}
\end{align}
with $\tsn(\SigmaISM,\Mth)$ given by equation (\ref{eq:tob_sigma}) with $\Mc = \Mth$. This expression can be solved numerically to obtain $\Mth$ as a function of only the ISM surface density $\SigmaISM$.

\begin{figure}
	\includegraphics[width=1.0\columnwidth]{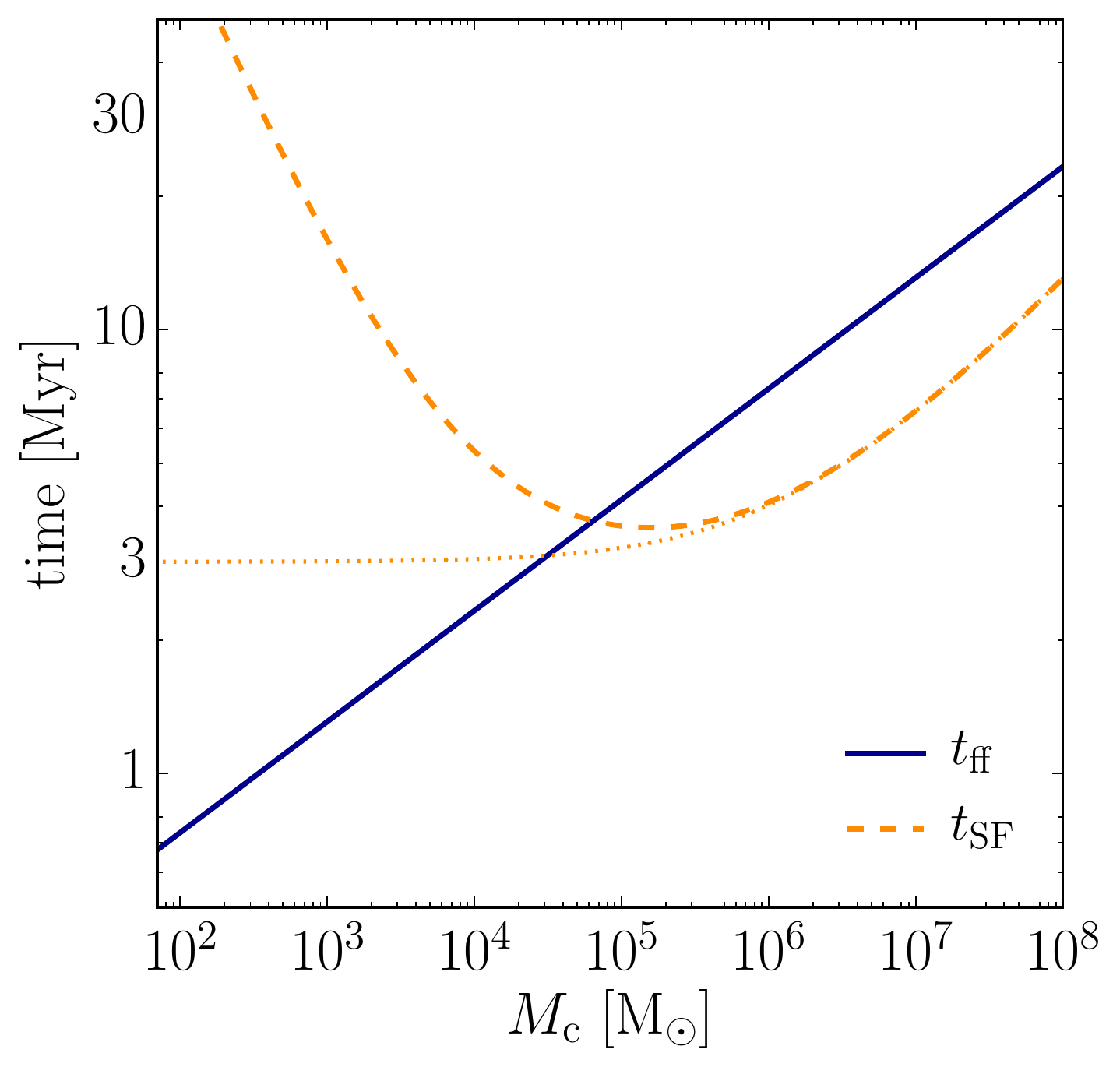}%
    \caption{Predicted star formation time-scales as a function of cloud mass in the environment of the solar neighbourhood. The solid and dashed lines show the free-fall time and the time until SN feedback stops star formation, respectively.  Clouds in the regime where $\tfb \gg \tff$ will form stars for several free-fall times and may reach a star formation efficiency high enough to remain bound after the remaining gas is expelled. Clouds with $\tfb < \tff$ will form stars for less than a free-fall time, becoming unbound after gas expulsion. The dotted line shows the effect of ignoring the time delay until the formation of a massive star introduced by the sampling of the IMF.}
    \label{fig:time-scales}
\end{figure}

To visualise the variation of the star formation time-scale with cloud mass, we show in Figure~\ref{fig:time-scales} the free-fall and feedback time-scales. As an example, we choose here the fiducial case of the solar neighbourhood environment and we assume an ISM surface density of $\SigmaISM = 13\Msun\pc^{-2}$ \citep{KennicuttEvans12}. Figure~\ref{fig:time-scales} shows the emergence of two regimes in the behaviour of the star formation time-scale. For large cloud masses, $\tfb$ increases with mass because the feedback energy per unit cloud mass decreases (see discussion of equation~\ref{eq:t_fb} in Section \ref{sec:minmass}), requiring that star formation proceeds for longer so that stellar feedback can match the inward pressure. Towards the regime of low cloud masses, the star formation time-scale first begins to saturate near its minimum value of $\tfb \sim 3 \Myr$ (the SN delay of the most massive OB star) and then rises again with decreasing mass due to the growing delay until the formation of the first massive star. 

\begin{figure}
	\includegraphics[width=1.02\columnwidth]{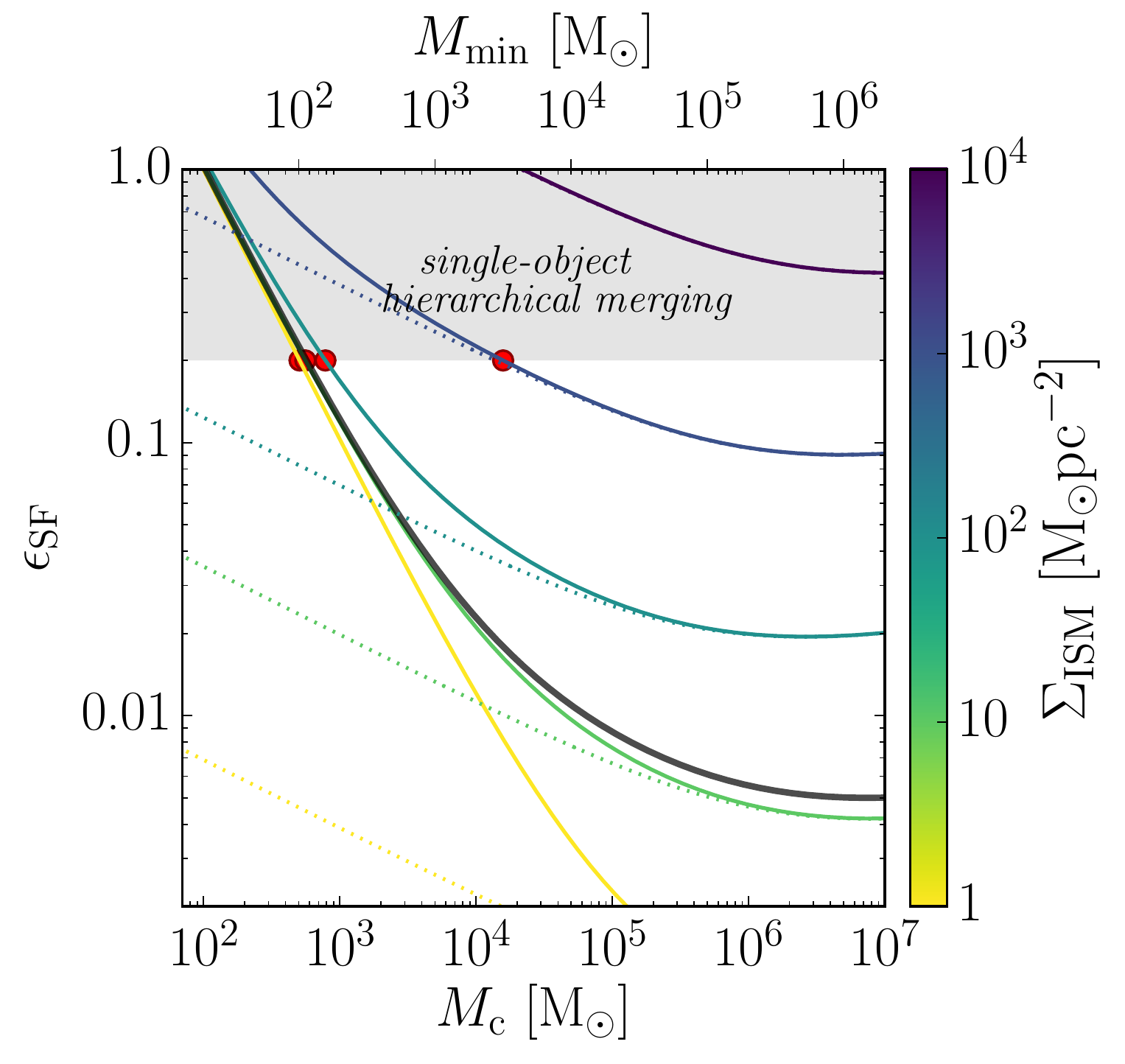}%
    \caption{Dependence of the feedback-regulated integrated star formation efficiency, $\epsfb$, on the cloud mass, $\Mc$, and ISM surface density, $\SigmaISM$. Clouds in the regime $\epsfb > 0.2$ convert gas into stars so efficiently that they are guaranteed collapse into a single bound object prior to residual gas expulsion. The circles mark the threshold cloud mass below which all stars must collapse into a single bound object, $\Mth$. The stellar mass of this object then becomes the bottom of the hierarchy of subclusters that populate the ICMF. The minimum mass of a bound stellar cluster is then $\Mmin = \epsfb \Mth$ and its values are indicated by the top axis. The dotted lines show the effect of neglecting the delay time until the formation of the first massive star on the resulting minimum mass. The thick black line indicates the prediction for the conditions in the solar neighbourhood.}
    \label{fig:boundcond_mass}
\end{figure}

The dependence of the condition to form a bound cluster on the ISM surface density is shown in Figure~\ref{fig:boundcond_mass}. The figure shows the total feedback-regulated star formation efficiency, $\epsfb$, as a function of cloud mass for $\SigmaISM = [1,10,10^2,10^3,10^4]\Msun\pc^{-2}$. At a given ISM surface density, when $\epsfb$ crosses into the single-object collapse region, all cloud masses below this threshold mass will collapse into a single bound cluster. The largest cloud mass scale $\Mth$, which limits where the stars are guaranteed to collapse into a single bound object, increases with host galaxy ISM surface density. For $\SigmaISM < 10\Msunpc2$, this mass is just below $10^3 \Msun$ and it increases rapidly to more than $10^5\Msun$ for $\SigmaISM > 10^{3.3}\Msunpc2$, following an asymptotic dependence of approximately $\Mth\propto\SigmaISM^{3}$. These threshold cloud mass values are marked by filled circles for each line corresponding to a fixed ISM surface density in Figure~\ref{fig:boundcond_mass}. For very large ISM surface densities, $\SigmaISM > 10^{3.5}\Msunpc2$, all cloud scales form bound clusters.

Due to the hierarchical nature of molecular cloud structure, all scales below the threshold bound mass will remain bound and will eventually merge. This implies that for a given host galaxy environment, the minimum mass of bound stellar clusters in our model is given by
\begin{equation}
    \Mmin(\SigmaISM) = \epsbound \epscore \Mth \simeq 0.2\Mth .
    \label{eq:Mmin_Mth}
\end{equation}
The minimum cluster mass is indicated in Figure~\ref{fig:boundcond_mass} for each $\SigmaISM$ using filled circles, with numerical values displayed along the top axis. The absolute minimum limit on the minimum cluster mass is set by $\MOB$ (see Section \ref{sec:IMF}). This can be understood by examining the behaviour of equation (\ref{eq:boundcondISM}) when $\Mth \downarrow 0$, which results in the limiting condition $\epsbound\epscore\Mth \geq \MOB$. 

In the regime of very large ISM surface densities, $\SigmaISM > 10^{3.5}\Msunpc2$, all cloud scales merge hierarchically into a single bound cluster, and this process is limited only by the fraction of the Toomre mass that is able to collapse under the influence of feedback (i.e.~the maximum cluster mass predicted by the \citetalias{Reina-Campos17} model). In this regime we set $\Mmin = \Mmax$, where $\Mmax$ is corrected relative to its classical form in \citetalias{Reina-Campos17} to account for the effect of IMF sampling (see Appendix~\ref{sec:appendix}). The maximum cluster mass can be expressed as a function of the ISM surface density $\SigmaISM$, the disc angular rotation velocity $\Omega$, and the \citet{Toomre64} stability parameter defined as
\begin{equation}
    Q \equiv \frac{ \kappa \vdispISM }{ \pi G \SigmaISM } ,
    \label{eq:Q}
\end{equation}
where $\kappa$ is the epicyclic frequency, 
\begin{equation}
    \kappa \equiv \sqrt{2} \frac{V}{R}\sqrt{1 + \frac{{\rm d} \ln V}{{\rm d} \ln R}} = \sqrt{2}\Omega ,
\end{equation}
where $V$ is the circular velocity at a galactocentric radius $R$, and the last equality holds for a flat rotation curve. This model thus introduces a dependence of the minimum mass on disc angular velocity and $Q$ in the regime of very high ISM surface density.

\section{Model uncertainties from empirically-derived parameters}
\label{sec:uncertainties}

\subsection{Star formation efficiency}

Our model for the environmental dependence of the minimum cluster mass hinges on the feedback-regulated total star formation efficiency, $\epsfb$, because it determines the maximum cloud mass ($\Mth$) and associated cluster mass ($\Mmin$) below which the star formation efficiency is so high that the young stars must collapse into a larger bound part of the hierarchy before the gas is lost. The main sources of uncertainty in $\Mmin$ are thus the empirically derived star formation efficiency per free-fall time, $\epsff$, and the product $\epsbound\epscore$.

The star formation efficiency per free-fall time of the cloud enters in equation~(\ref{eq:boundcondISM}) with a scaling $\epsfb \propto \epsff$ (assuming for simplicity that $\tsn = \tsnO$ and neglecting the subdominant second term inside the square root). Many estimates of $\epsff$ are available in the literature. For instance, \citet{Utomo18} obtain the largest and most direct sample of measurements of $\epsff$ in external galaxies using CO observations at the scales of typical GMCs. The authors find a mean value $\epsff = 0.7 \pm 0.3$ per cent across the sample of 14 galaxies. However, larger values, $\epsff = 1.5 - 2.5$ per cent, are found in studies of individual Milky Way clouds \citep{Evans14, Lee16, Barnes17}, while lower values, $0.3 - 0.36$ per cent, are observed in M51 \citep{Leroy17}. Taken together, these results imply an uncertainty in the efficiency per free-fall time of at least a factor of $\sim 2$ in each direction. 

Beyond the uncertainty in the determination of $\epsff$ in local galaxies, there is also the additional possibility that the star formation efficiency $\epsff$ is not universally constant, but instead varies as a function of time and cloud properties. For instance, in numerical simulations of isolated molecular clouds including stellar feedback, \citet{Grudic18} find that $\epsff$ correlates with cloud surface density. Observational studies where star-forming regions in the Milky Way are matched to the nearest molecular clouds find a much larger scatter in the star formation efficiency $\sigma_{\log \epsff} \sim 1$ \citep{Evans14,Lee16,Vutisalchavakul16} compared to all other methods, including YSO counting and pixel statistics. \citet{Lee16} explain this large scatter using a model with a strong time-dependence of the star formation efficiency. However, \citet{Ochsendorf17} find that, when applied to the same data set, the cloud matching method tends to overestimate the median and the scatter in $\epsff$ compared to the YSO counting method because it associates the entire flux from a star-forming region to its nearest GMC, even in regions where there is no overlap or physical association. Indeed, different cloud matching studies using the same sample obtain mean efficiencies that are different by a factor of almost $\sim 10$ due to a difference in the sensitivity of the observations \citep{Krumholz19}. 
This bias in the cloud matching studies is also consistent with the expectation that star formation relations observed at galactic scales break down below the spatial and temporal scales of individual clouds due to statistical under-sampling of independent regions \citep{KruijssenLongmore14, Kruijssen18a}.  

Following the recent review of observational measurements of $\epsff$ by \citet{Krumholz19}, we adopt the constant value $\log \epsff = -2$ with a systematic uncertainty $\sim 0.5$ dex, which is consistent with all other methods except cloud matching. Implementing a time dependent $\epsff$ would require following the time evolution of the structure of the gas within clouds, and this is beyond the scope of the simple model presented here. For future work, it would be interesting to explore of the effect of a dependence on cloud surface density \citep{Grudic18}. At present, the only measurement of $\epsff$ in the high-surface density regime (in the Central Molecular Zone of the MW) yields $\epsff \approx 1.8$ per cent \citep{Barnes17}, consistent with the values found for lower surface density clouds in the solar neighbourhood \citep[e.g.][]{Evans14, Heyer16, Ochsendorf17}.

\begin{figure*}
	\includegraphics[width=1.0\columnwidth]{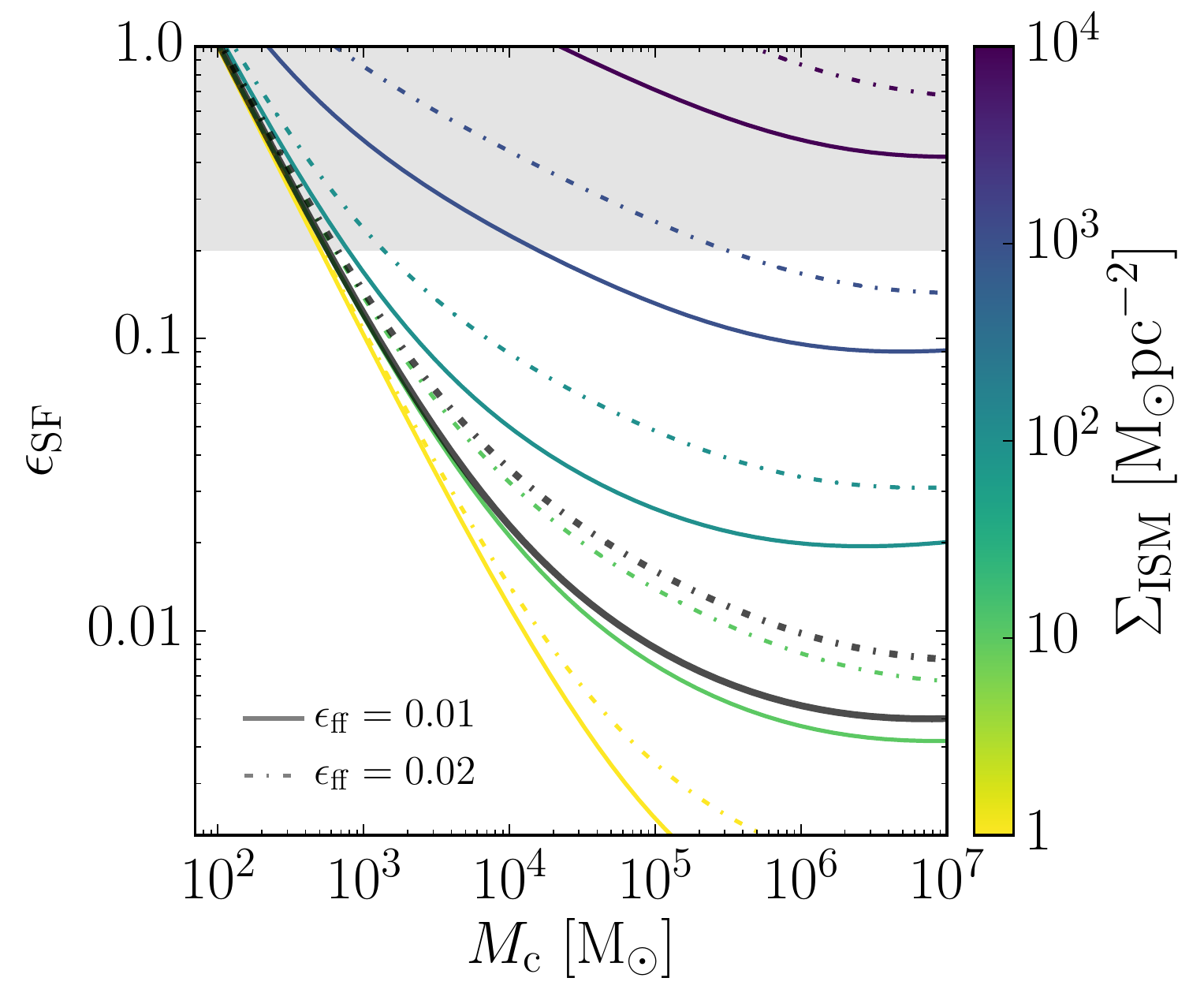}%
	\includegraphics[width=1.0\columnwidth]{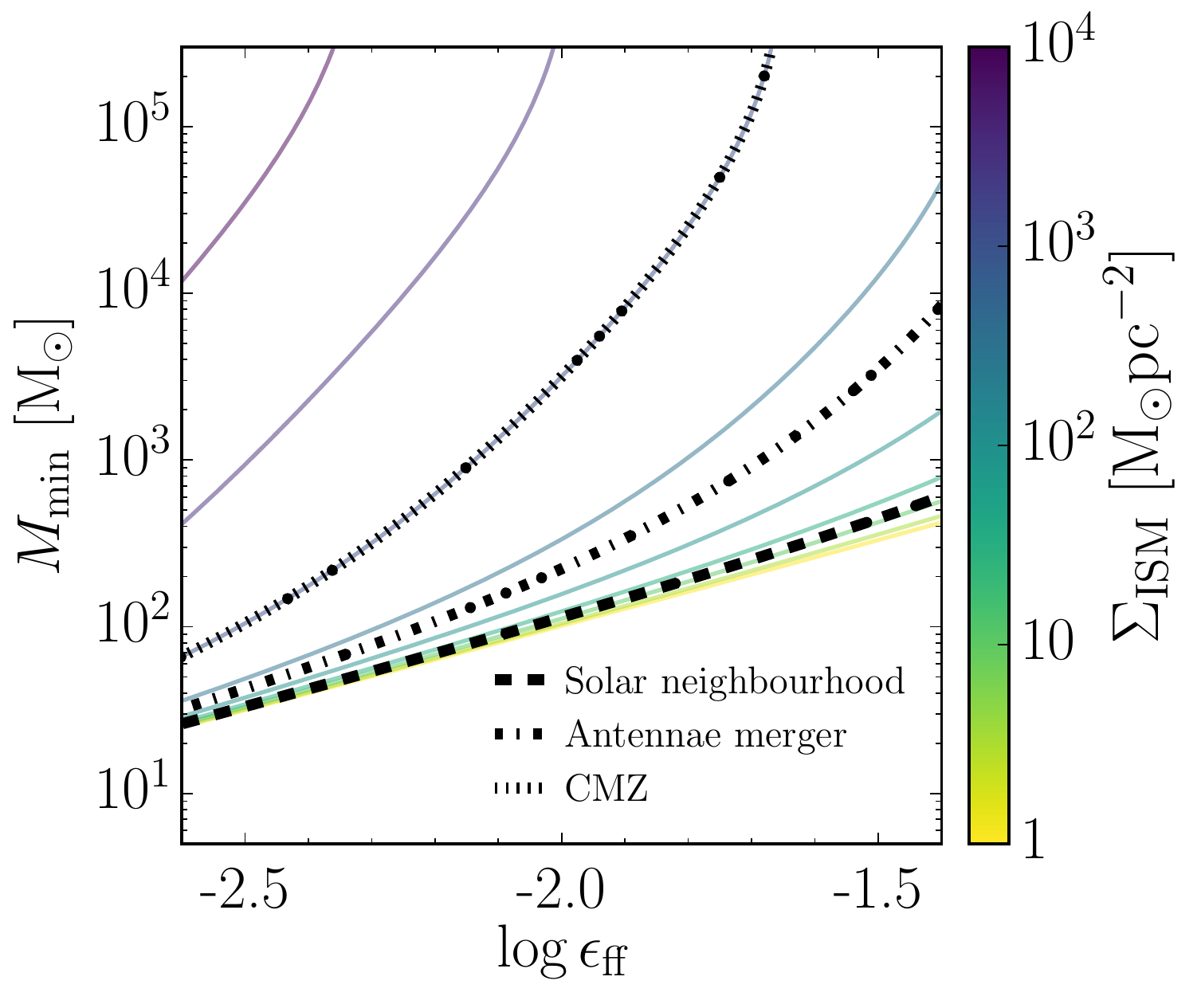}%
    \caption{Impact of the value of star formation efficiency per free-fall time ($\epsff$) on the predicted minimum cluster mass. {\it Left panel}: reproduction of Figure~\ref{fig:boundcond_mass} with values of $\epsff = 1$ per cent (fiducial model; solid lines), and $\epsff = 2$ per cent (dot-dashed lines). The threshold cloud mass (where $\epsfb = \epsbound\epscore$ and the lines cross into the shaded region) is very sensitive to the assumed value of $\epsff$ for ISM surface densities $\SigmaISM \ga 10^2\Msunpc2$, but saturates at lower surface densities due to the delay in feedback caused by IMF sampling in low mass clouds. The thick black line indicates the prediction for the solar neighbourhood. {\it Right panel}: shows the dependence of the minimum cluster mass on $\epsff$  for lines of constant ISM surface densities. The solid, dashed, and dot-dashed black lines indicate the surface gas densities in the solar neighbourhood, the Antennae galaxy merger, and the CMZ, respectively.}
    \label{fig:boundcond_mass_eps_ff}
\end{figure*}

According to equation~(\ref{eq:boundcondISM}), and neglecting again the second term inside the square root, at a fixed ISM surface density in the regime $\SigmaISM \ga 10^2\Msun$ the total star formation efficiency scales approximately as $\propto \Mc^{-1/4}$ (i.e.~the region below the turnover for each of the dotted lines of Figure~\ref{fig:boundcond_mass}). This implies that, for a fixed value of $\epsbound\epscore$, a factor of two change in $\epsfb$ results in a factor of $\sim 2^4 = 16$ difference in the threshold cloud mass in the high ISM surface density regime ($\SigmaISM \ga 10^2\Msun$). This scaling requires that the effect of IMF sampling is minor, which means the dependence of $\Mmin$ on $\epsfb$ is largest for ISM surface densities $\SigmaISM \ga 10^2\Msunpc2$. The sensitivity of the total star formation efficiency $\epsfb$ to an uncertainty of a factor of two in $\epsff$ is illustrated in the left panel of Figure~\ref{fig:boundcond_mass_eps_ff}. Moreover, the panel also shows that the minimum mass in low ISM surface density environments, $\SigmaISM \la 10^2\Msunpc2$ \citep[such as in local disc galaxies;][]{Kennicutt98}, is much more robust as its scaling with cloud mass is much steeper (approximately $\propto \Mc^{-1}$) due to the effect of IMF sampling discussed in Section~\ref{sec:IMF}. The right panel of Figure~\ref{fig:boundcond_mass_eps_ff} shows the explicit dependence of the minimum cluster mass on $\epsff$ for lines of constant ISM surface density, indicating the conditions across various galactic environments including the solar neighbourhood, the Antennae galaxies, and the CMZ (see Section~\ref{sec:environments} for a description of the parameters used). The minimum cluster mass is a sensitive probe of the physics of star formation in high surface density environments like the CMZ. 

Because the value of $\epsbound$ is reasonably well constrained by simulations, the uncertainty in the product $\epsbound\epscore$ is dominated by the limiting efficiency in molecular cores, $\epscore$. Also known as the core-to-star efficiency, its value is constrained by observations, analytical models and simulations to the range $\epscore \simeq 0.3 - 0.7$ \citep{MatznerMcKee00, Enoch08, FederrathKlessen12, FederrathKlessen13}. This relative uncertainty is therefore about a factor of 2 smaller than the $\epsff$ uncertainty and its effect on the minimum mass is in the opposite direction. Assuming $\epscore > 0.5$ lowers $\Mmin$ for a fixed $\epsff$, while assuming $\epsff > 0.01$ increases $\Mmin$ by the same amount for a fixed $\epscore$. Increasing both parameters by a factor of 2 would result in an overall change in $\Mmin$ of less that a factor of 2. 

To summarise, we highlight the following points:
\begin{enumerate}
    \item The minimum cluster mass predicted by our model for \emph{high ISM surface density environments} ($\SigmaISM \ga 10^2\Msunpc2$) is very sensitive to the assumed star formation efficiency per free-fall time, $\epsff$, and gas conversion efficiency in pre-stellar cores, $\epscore$. More stringent constraints on these values from future observations will be necessary to make more precise minimum cluster mass predictions in environments with ISM surface densities $\SigmaISM \ga 10^2\Msunpc2$. 
    \item Despite the sensitivity of the minimum mass on $\epsff$, the relative \emph{scaling} of the minimum cluster mass with ISM surface density that was obtained in the previous section (i.e.~the slope of the lines in Figure~\ref{fig:boundcond_mass_eps_ff}) is a robust prediction of our model in the regimes where the ISM surface density $\SigmaISM \ll 10^2\Msunpc2$ (e.g.~quiescent discs) and $\SigmaISM \gg 10^2\Msunpc2$ (e.g.~galactic nuclei).
    \item A benefit of the sensitivity of the minimum cluster mass to the star formation efficiency per free-fall time $\epsff$ in the high surface density regime is that $\Mmin$ is an independent observational probe of the small-scale physics of star formation.
\end{enumerate}

\subsection{The effect of radiation and stellar winds on the feedback timescale}
\label{sec:earlyfeedback}

The model presented here assumes that the feedback energy injection begins after the onset of the first supernova, $\tsn = 3\Myr$ (see equation \ref{eq:t_fb_def}). However, stellar winds and radiation could have a significant role in cloud disruption due to their nearly instantaneous onset compared to the delayed effect of SNe.  As a result of this, the relative role of each of these processes in regulating star formation is still highly debated in the literature \citep[for a review, see][]{Krumholz19}. To determine the sensitivity of the ICMF model to the effect of early feedback from radiation and stellar winds, we must consider both the total momentum injected, and the timescale over which it stops star formation. 

The momentum injection rates from stellar winds, radiation, and SNe are all comparable \protect\citep{Agertz13}. As shown in Figure~\ref{fig:time-scales}, the feedback timescale at low cloud masses ($\la 10^5\Msun$ in the solar neighbourhood) is dominated by the delay in the formation of the first massive star due to IMF sampling. In the majority of parameter space populated by galaxies, this sets the threshold cloud mass below which all scales remain bound. Indeed, neglecting the IMF delay, these clouds shut off their star formation very quickly after the onset of feedback, with $\tfb \sim \tsn$ (dotted line in Figure~\ref{fig:time-scales}). Increasing the energy injection parameter $\phifb$ would have a negligible effect on the duration of star formation because feedback is already extremely efficient in clouds with low enough masses to be affected by IMF sampling (the left branch of the $\tfb$ curve in Figure~\ref{fig:time-scales}).
 
\begin{figure}
	\includegraphics[width=1.0\columnwidth]{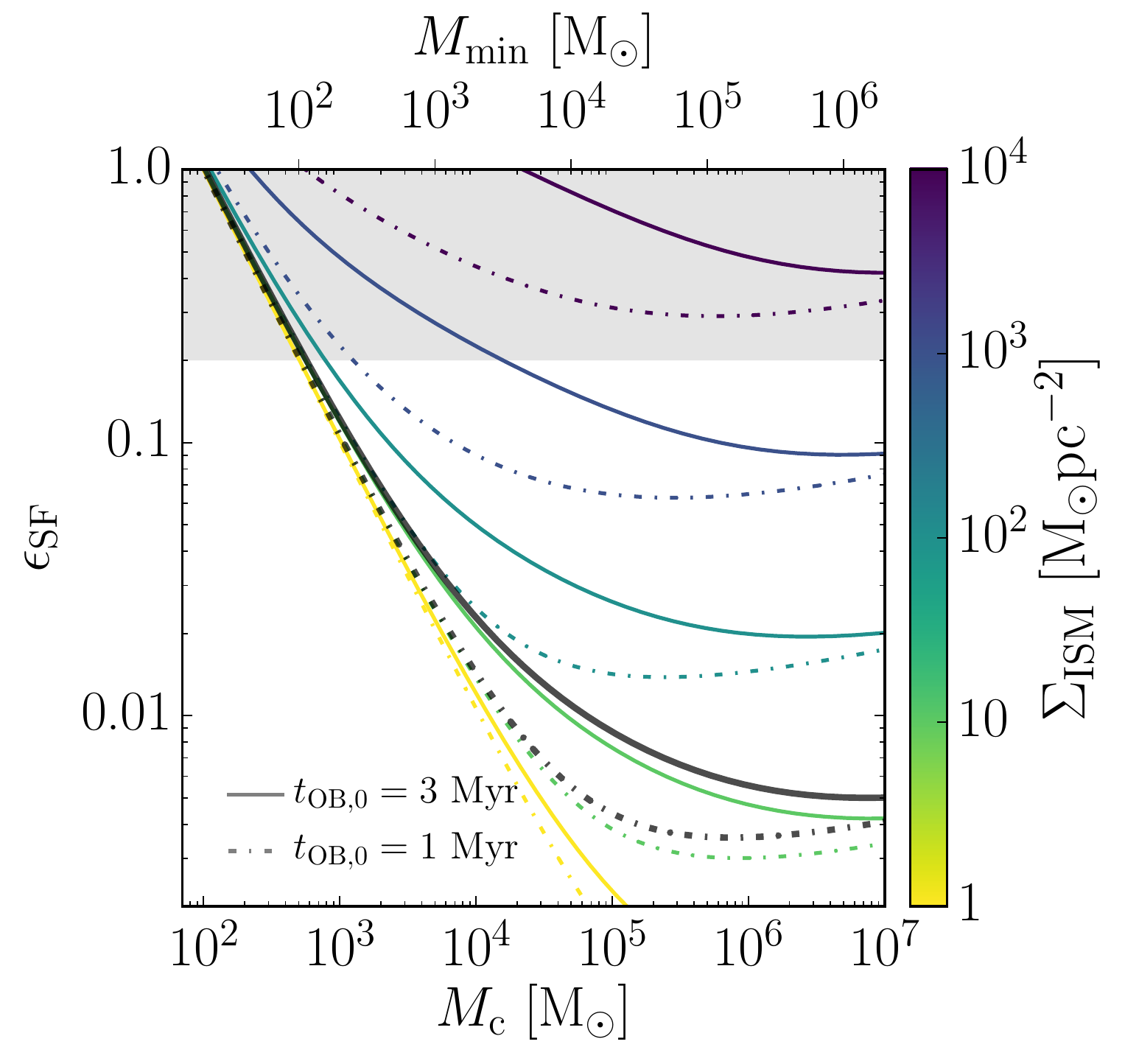}%
    \caption{Impact of the timescale for the onset of feedback, $\tsnO$ on the predicted minimum cluster mass. We show a reproduction of Figure~\ref{fig:boundcond_mass} with values of $\tsnO = 3\Myr$ (fiducial model; solid lines), and $\tsnO = 1\Myr$ (dot-dashed lines). The threshold cloud mass (where $\epsfb = \epsbound\epscore$ and the lines cross into the shaded region) is sensitive to the assumed value of $\tsnO$ only for high ISM surface densities, $\SigmaISM \ga 10^2\Msunpc2$. At lower surface densities the efficiency is driven mainly by the feedback onset delay caused by IMF sampling in low mass clouds. The thick black line shows the prediction in the solar neighbourhood.} 
    \label{fig:boundcond_mass_tsn0}
\end{figure}

The termination of star formation in simulated clouds can occur before the first SNe explode, $\tfb < \tsn$, when radiative feedback is included \protect\citep{Grudic18}. This is also expected from simplified analytical arguments \citep{Murray10}. However, it is difficult to derive a single timescale for the termination of star formation by radiation and stellar winds due to their highly nonlinear dependence on cloud structure. For massive clusters, all of these mechanisms become important \citep{Krumholz19}. It is also challenging to constrain the importance of radiation and stellar winds using observations of individual clouds because their entire evolutionary sequence is not observable. However, methods that rely on the statistics of star formation and molecular gas tracers across entire galaxies can constrain the mean evolutionary timescales of clouds \citep{KruijssenLongmore14, Kruijssen18a}. Using this approach, \citet{Kruijssen19b} and \citet{Chevance19} obtain the typical duration of star formation across several nearby spirals, $t_{\rm SF} - \tsn \sim 3 \Myr$. Such a short timescale implies that early feedback from radiation and stellar winds, and not SNe, regulate the star formation process in the conditions typically found in the local universe. In these conditions, our model predicts that molecular clouds with masses $\Mc \la 10^6\Msun$ will have a feedback stage duration $t_{\rm fb} = 3-4\Myr$ after forming a massive star. This is in broad agreement with the observed value.

In summary, two factors limit the effect of early feedback on the minimum cluster mass. First, the energy injection rate due to SNe alone effectively halts star formation on a very short timescale at the low cloud masses which define the bottom of the cluster hierarchy. This makes the model insensitive to the increase in the energy injection due to stellar winds and radiation. Second, the observed feedback timescale is $\sim 3\Myr$ in nearby spirals, in agreement with the total duration of star formation at the cloud masses that set the minimum cluster mass. 

To illustrate the effect of assuming a shorter feedback timescale, Figure~\ref{fig:boundcond_mass_tsn0} shows the integrated star formation efficiency and minimum cluster mass for $\tsnO = 1\Myr$. A factor of three reduction in the feedback onset time  results in negligible change in the minimum cluster mass for gas surface densities $\SigmaISM \la 10^2\Msunpc2$ due to the dominance of the feedback delay due to IMF sampling. At larger surface densities early feedback reduces the integrated star formation efficiency and the resulting minimum mass by up to a factor of $\sim 10$. However, at large surface densities, SNe, direct radiation pressure, and ionisation become less effective \citep{Krumholz19}, and this effect could increase the star formation efficiency in this regime. Although improved constraints on the feedback timescale will reduce this uncertainty in the future, Figure~\ref{fig:boundcond_mass_tsn0} shows that the results will not change qualitatively.

\section{The minimum cluster mass across the observed range of galactic environments}
\label{sec:environments}

The model described in Section~\ref{sec:model} defines the ICMF as a function of three parameters of the host galaxy. The low-mass truncation $\Mmin$ is determined by the gas surface density $\SigmaISM$ using equations  (\ref{eq:tob_sigma})-(\ref{eq:Q}), and the high-mass truncation $\Mmax$ is given by $\SigmaISM$, the angular rotation velocity of the disc $\Omega$ (or the epicyclic frequency), and Toomre $Q$ using equations (\ref{eq:tfb_Mmax}) - (\ref{eq:Mmax}) (see discussion below). In this section, we explore the behaviour of the ICMF truncation masses, $\Mmin$ and $\Mmax$, within the three-dimensional parameter space spanned by these parameters.  

\begin{figure*}
    \includegraphics[width=\textwidth]{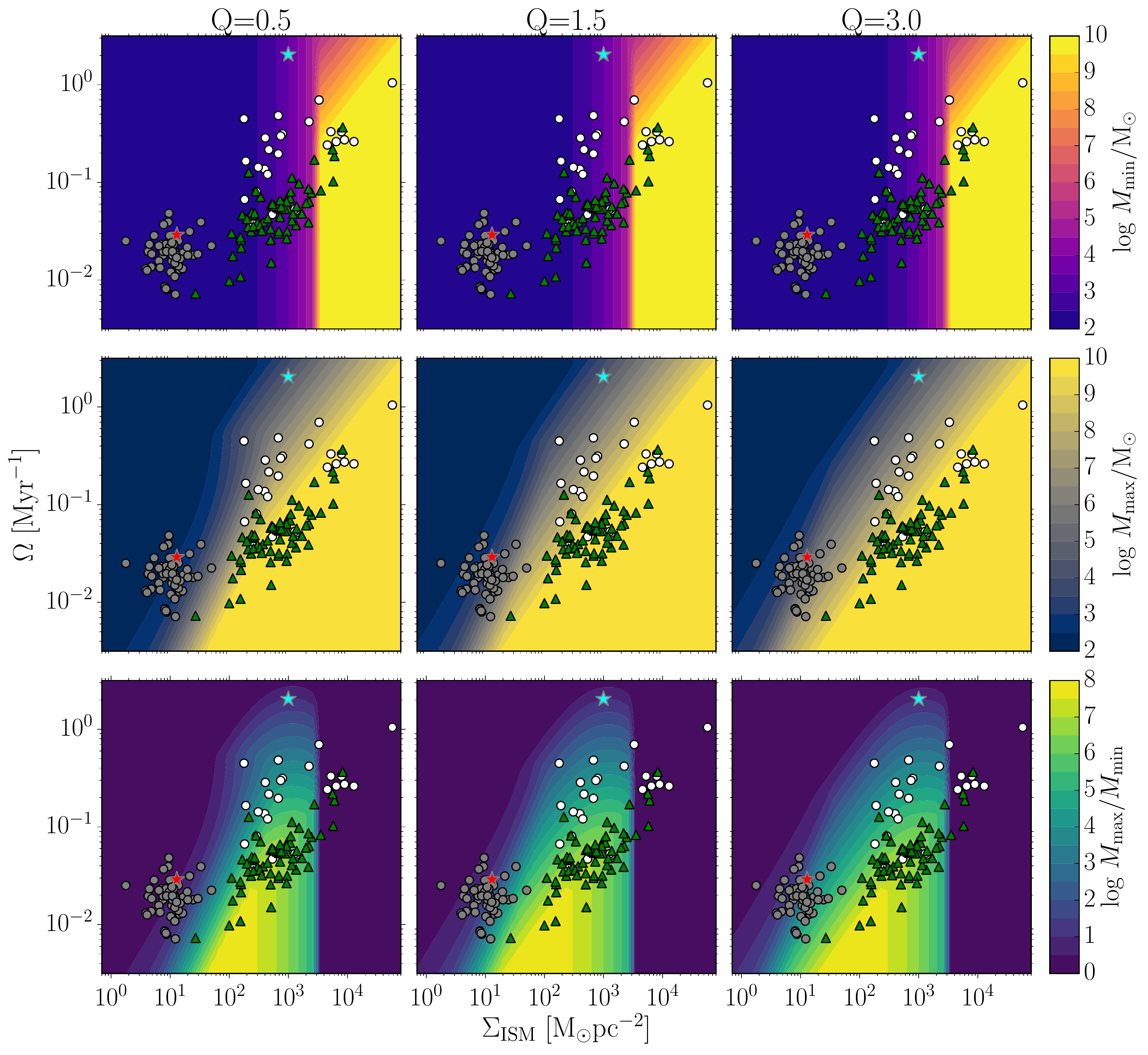}%
    \caption{Minimum and maximum masses of bound star clusters as a function of the galactic environment, spanned by the ISM surface density ($\SigmaISM$) and the angular velocity of the rotation curve ($\Omega$). From left to right, columns show values of the Toomre parameter $Q \in [0.5, 1.5, 3.0]$. \emph{Top row:} the minimum bound cluster mass. \emph{Middle row:} the maximum cluster mass obtained using the model of \citetalias{Reina-Campos17} modified to account for IMF sampling in low-mass clouds. \emph{Bottom row:} logarithmic width of the ICMF, defined as the logarithmic of the ratio between the maximum and minimum cluster mass. The locations of observed galaxies within the parameter space are indicated by the symbols in each panel. Grey and white circles represent the star-forming galaxies and starbursts from \citet{Kennicutt98} respectively, while green triangles are the high-redshift galaxies from \citet{Tacconi13}. The red star corresponds to the solar neighbourhood, and the cyan star to the CMZ of the Milky Way. The surface density of the ISM is the main driver of the variation in $\Mmin$, with the minimum mass remaining almost constant for ISM surface densities $\SigmaISM \la 10^2\Msun\pc^{-2}$ and then steeply rising as $\sim \SigmaISM^3$ for larger values. The different environmental dependence of the minimum and maximum mass scales leads to a wide variety of predicted ICMF widths for observed galaxies.}
    \label{fig:paramspace}
\end{figure*}

The top row of Figure~\ref{fig:paramspace} shows the dependence of the minimum cluster mass on the ISM surface density and angular velocity (obtained from solving equation~\ref{eq:boundcondISM}). The columns show, from left to right, the results for three different Toomre parameter values $Q \in [0.5, 1.5, 3.0]$. The range of the colorbar has an upper limit at $M=10^{10}\Msun$ for clarity. The symbols reproduced in each panel represent observations of star-forming galaxies and starbursts from \citet{Kennicutt98}, high-redshift galaxies from \citet{Tacconi13}, the solar neighbourhood, and the Milky Way's CMZ. For the solar neighbourhood, we consider an ISM surface density $\SigmaISM = 13\Msun\pc^{-2}$ and $\Omega = 0.029\Myr^{-1}$ (see Section~\ref{sec:global}). For the CMZ, we use $\SigmaISM \sim 10^3\Msun\pc^{-2}$ \citep{Henshaw16} and calculate the angular velocity using the enclosed mass profile from \citet{Kruijssen15a}. The minimum cluster mass depends mainly on the ISM surface density across most of the parameter space occupied by galaxies. The dependence on the angular velocity only becomes significant in the top right region, mostly corresponding to galactic nuclei, where both the ISM surface density and angular velocity are high. This is the region where $\Mmin = \Mmax$, because the entire cloud hierarchy merges into a single bound object. There is a large variation of the minimum mass with galactic environment, with the sequence of observed galaxies, from local discs to high-redshift galaxies, spanning $\sim 5$ orders of magnitude in minimum stellar cluster mass.

The physical mechanisms setting the minimum cluster mass also vary with the galactic environment. For low ISM surface densities ($\SigmaISM \la 10^2\Msunpc2$), the nearly constant minimum mass $\Mmin \sim 10^{2-2.5}\Msun$ is caused by the delay in the formation of the first massive star setting a fixed lower limit to the cloud mass that produces a bound cluster (see Figure~\ref{fig:boundcond_mass}). For ISM surface densities typical of high-redshift galaxies ($\SigmaISM > 10^2\Msunpc2$), the minimum bound cluster mass scales with the mean ISM surface density approximately as $\Mmin \propto \SigmaISM^3$. Physically, this corresponds to the regime dominated by the steep dependence of the free-fall time on the ISM surface density ($\epsfb \propto \tff^{-1} \propto \SigmaISM^{3/4}/\Mc^{1/4}$) in the first term on the right-hand side of equation~(\ref{eq:boundcondISM}). As the ISM surface density increases, increasingly massive clouds ($\Mc \propto \SigmaISM^3$) achieve such high star formation efficiencies ($\epsfb$) that they must collapse into a single bound cluster. 

At very large ISM surface densities ($\SigmaISM \ga 2\times10^3\Msunpc2$) the scaling becomes increasingly steeper because the second term inside the square root in equation (\ref{eq:boundcondISM}) becomes important, causing the minimum of the $\epsfb$ versus $\Mcloud$ curve (see Figure~\ref{fig:boundcond_mass}) to approach the threshold value $\epsbound\epscore$ at an increasing rate. This is the regime where stellar feedback becomes increasingly inefficient and the feedback timescale grows with cloud mass, allowing for higher integrated star formation efficiencies ($\epsfb$). For the largest observed surface densities ($\SigmaISM > 4\times10^3\Msunpc2$), the minimum of the $\epsfb$ curve is entirely above the threshold efficiency (see Figure~\ref{fig:boundcond_mass}), and all cloud masses will collapse into a single bound object limited in mass only by the maximum cluster mass. This mass is determined by the collapse of the largest unstable scale in the \citetalias{Reina-Campos17} model (see Section~\ref{sec:global}).  

As can be seen in Figure~\ref{fig:paramspace}, the minimum cluster mass has a negligible dependence on the Toomre parameter. This occurs in the high-ISM surface density and angular velocity regime and originates from the behaviour of the maximum mass. As a result of this shift, most observed galaxies with low shear (low angular velocity) have a minimum cluster mass that depends only on the ISM surface density.

The middle row of Figure~\ref{fig:paramspace} shows the maximum cluster mass predicted by the \citetalias{Reina-Campos17} model. In order to be consistent with the minimum cluster mass model presented in Section~\ref{sec:model}, we have updated the original \citetalias{Reina-Campos17} model to include the effect of IMF sampling. The modifications are described in Appendix~\ref{sec:appendix}. As discussed in Section~\ref{sec:global}, at gas surface densities above $\sim 4\times 10^3\Msunpc2$ the minimum and maximum cluster mass are equal because the entire cloud mass spectrum will collapse into a single bound object at the maximum mass scale.

We combine the minimum mass model with the model for the maximum cluster mass from \citetalias{Reina-Campos17} to predict the full width of the ICMF and its dependence on the galactic environment. The bottom panels of Figure~\ref{fig:paramspace} show the logarithmic width of the ICMF as a function of ISM surface density and angular velocity for values of $Q \in [0.5, 1.5, 3.0]$.

Because of the intrinsically different dependence of the minimum and maximum mass scales on the ISM surface density and angular velocity, the predicted width of the ICMF shows a large non-monotonic variation within the region of parameter space populated by observed galaxies. The model predicts that galaxies with gas surface densities, $10 \la \SigmaISM \la 100\Msunpc2$ and slow rotation, $\Omega \la 0.03\Myr^{-1}$, such as local quiescent discs, will have relatively broad mass functions. On the other extreme, galactic environments with either high gas ISM surface densities, $\SigmaISM \ga 2\times 10^3\Msunpc2$ (e.g.~massive high-redshift discs), or fast rotation, $\Omega \ga 0.5\Myr^{-1}$ (e.g.~galactic nuclei) should have narrow ICMFs.

\section{Comparison to observed cluster mass functions in the local Universe}
\label{sec:predictions}

After exploring the general predictions of our minimum cluster mass model for the range of observed galaxy properties, we turn our attention to the detailed predictions for the ICMF in nearby galaxies where observational constraints are currently available. As a result of observational systematics that are hard to correct for with current data, it is extremely challenging to determine observationally the abundance of low-mass ($\la 10^3\Msun$) clusters in nearby galaxies, including the Milky Way. None the less, current observations provide valuable upper limit constraints for the low-mass truncation of the ICMF predicted by the model. The predictions we provide here for $\Mmin$ and for the full ICMF should become testable with upcoming observational facilities, such as 30-m class telescopes, in the near future. Here we assume the model ICMF defined as a power law with index $\beta=-2$ and exponential truncations at the minimum ($\Mmin$,  equation~\ref{eq:boundcondISM}) and maximum ($\Mmax$, equation~10 of \citetalias{Reina-Campos17} modified to include the effects of IMF sampling; see Appendix~\ref{sec:appendix}) mass scales, as described in equation~(\ref{eq:CMF}).

Since the observed CMF evolves rapidly after several million years due to dynamical effects, which are not included in the model \citep[e.g.][]{BaumgardtMakino03,Lamers05,Kruijssen12c}, we restrict the comparison to the mass functions of observed young clusters in two separate age ranges, $\tau \la 10\Myr$ and $\tau \la 100\Myr$, where $\tau$ is the cluster age. In observational samples where these ranges are not available, we use the two lowest age bins. Clusters in the youngest bin should be least affected by dynamical disruption, retaining the initial mass distribution, while the older clusters should allow a more complete statistical sampling of the high mass tail of the ICMF (which is also the most insensitive to tidal disruption). However, the youngest age bin is also the most strongly affected by contamination by unbound associations \citep{bastian12,kruijssen16}

In the following sections, we compare the model predictions to observations of the young CMF. These are chosen to represent a broad range of star-forming conditions, including the solar neighbourhood and the LMC as examples of low-ISM surface density environments, and the Antennae galaxies and the Milky Way's CMZ representing conditions of high density and high shear.  

\subsection{The solar neighbourhood}
\label{sec:sn}

The Milky Way is an ideal place to test predictions for the low-mass turnover of the ICMF. The deepest limits can be obtained in the solar neighbourhood, such that the turnover of the CMF at the minimum mass might be detectable. 

\begin{figure*}
    \includegraphics[width=1.0\textwidth]{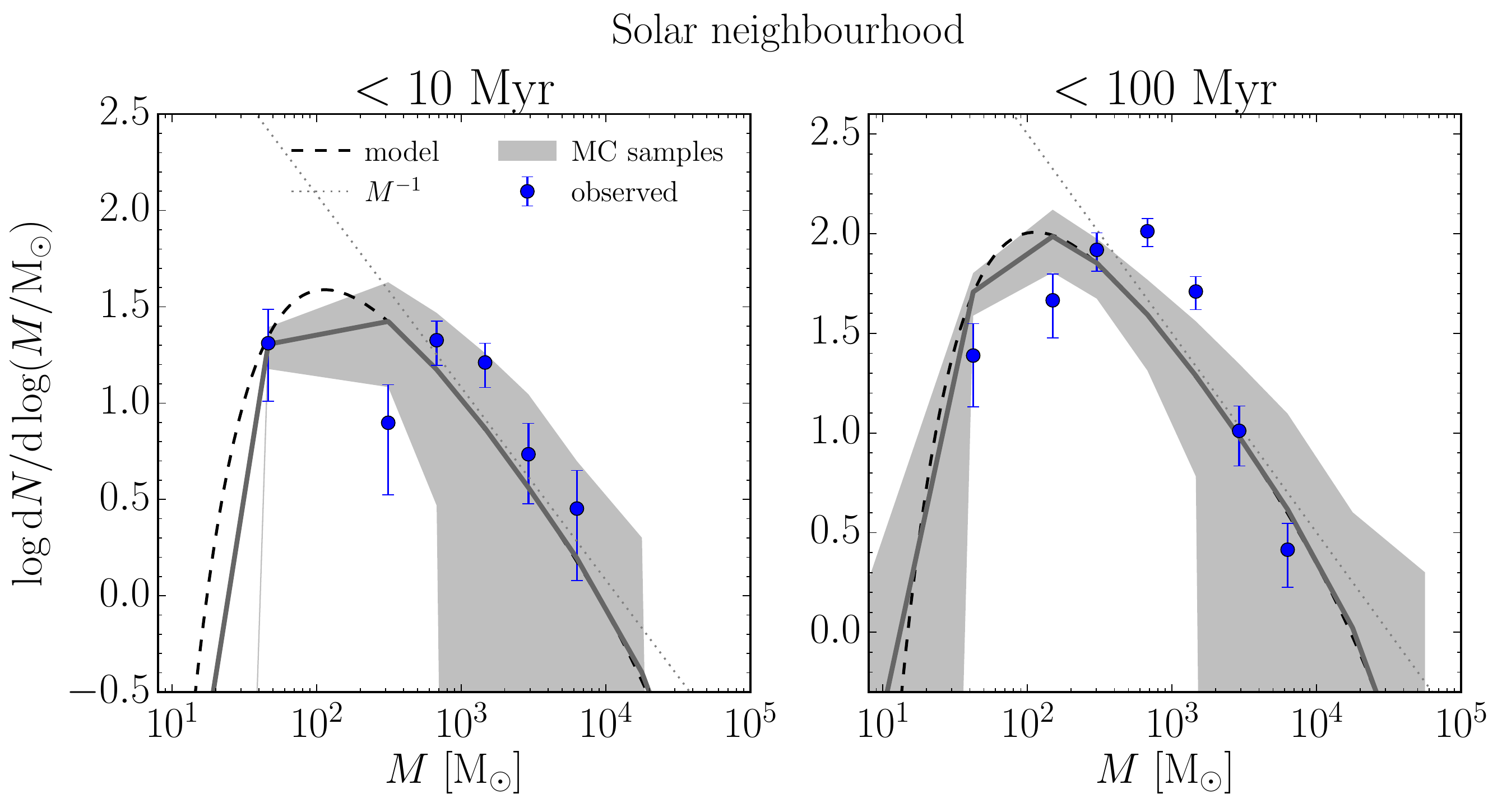}%
    \caption{Predicted and observed young CMF in the solar neighbourhood for clusters in two different age ranges (left: $\tau < 10\Myr$; right: $\tau <100\Myr$). The prediction of our ICMF model is shown as a dashed line, adopting the maximum mass from the \citetalias{Reina-Campos17} model and the functional form from equation~(\ref{eq:CMF}). The solid line and shaded band show the mean and the 2.5 to 97.5 percentile range of a Monte Carlo sampling of the predicted ICMF. The dotted line shows a pure power-law CMF with $\beta = -2$. The data points correspond to the cluster sample used by \citet{Lamers05}. Despite some deviations at low masses, the predicted minimum cluster mass is consistent with the observed turnover for clusters in the solar neighbourhood. }
\label{fig:CMF_MW}
\end{figure*}

To determine the observed CMF in the solar neighbourhood we use the \citet{Kharchenko05} cluster catalogue. The catalogue contains  homogeneously determined cluster membership, distances and apparent magnitudes. For cluster masses we use estimates from \citet[and H.~J.~G.~L.~M.~Lamers, private communication]{Lamers05}. We then calculate the CMF in two age bins following the procedure in \citet{Piskunov08}, with mass bins chosen to reduce sampling errors. We chose the normalisation of the theoretical ICMF to yield the same total number of clusters in the relevant mass range as the catalogue in the surveyed area. The result is shown in Figure~\ref{fig:CMF_MW}.

The main issues affecting the precise determination of the CMF in the Milky Way are incompleteness due to dust extinction for the faintest clusters as well as uncertainties in membership determination. In addition, many clusters lack mass estimates due to the low number of available member stars. These systematic effects are difficult to quantify, so the data for the lowest mass clusters should be interpreted with caution. The error bars we quote represent the statistical uncertainty and are therefore only lower limits on the total uncertainties.

To obtain our model prediction for the minimum cluster mass in the solar neighbourhood environment, we use the observed ISM surface density, $\SigmaISM =13\Msun~\pc^{-2}$ \citep{KennicuttEvans12}, and angular velocity (assuming a flat rotation curve), $\Omega =0.029\Myr^{-1}$ \citep{Bland-HawthornGerhard16}. Using these values and the typical observed ISM velocity dispersion, $\sigma_{\rm ISM} = 10\kms$ \citep{HeilesTroland03}, we derive the Toomre $Q$ parameter (equation~\ref{eq:Q}). 

Figure~\ref{fig:CMF_MW} shows a direct comparison of the observed CMF for clusters of ages $\tau < 10$ and $\tau < 100\Myr$ with the prediction of our model. To include the effects of discrete sampling in the tails of the distribution function, we also produce Monte Carlo samples by drawing clusters from the predicted mass distribution until the total number of clusters in the sample matches the total observed number. The model predicts a low-mass truncation of the ICMF at a minimum mass $\Mmin = 1.1\times10^2\Msun$, and a maximum cluster mass  $\Mmax = 2.8\times10^4\Msun$.

It is challenging to determine the completeness limit of Milky Way cluster catalogues. For this reason, the apparent low-mass turnover in the observed CMF in Figure~\ref{fig:CMF_MW} is hard to interpret, as it could be caused by incompleteness. At low cluster masses, significant systematic errors arise due to large extinction corrections and a lack of mass estimates because of the low number of stars detected in the faintest clusters. In fact, \citet{Cantat-Gaudin18} recently used \emph{Gaia} data to show that the current cluster catalogues in the solar neighbourhood could be highly incomplete. Taking into account the large uncertainties at the lowest masses, both age bins in the observed CMF agree well with our model in the regime $M \geq 10^3\Msun$, and are qualitatively consistent with the predicted turnover at low masses.

\subsection{M31}
\label{sec:m31}

M31 is by far the nearest massive extragalactic disc galaxy, making the determination of the CMF more straightforward and less prone to systematics than observations of the cluster population in the solar neighbourhood. To predict the ICMF for M31, we use the sum of the molecular and atomic gas surface densities, $\SigmaISM = 9.3\Msunpc2$ (A.~Schruba et al., in prep.), an angular velocity of $\Omega = 0.021\Myr^{-1}$, and a Toomre parameter of $Q = 2.1$. The angular velocity and Toomre parameter are derived for a galactocentric radius of $12\kpc$ corresponding to the star-forming ring where most of the cluster formation takes place. The rotation curve at this radius is approximately flat with a circular velocity $V_{\rm flat} = 250\kms$ \citep{Corbelli10}, and a  gas velocity dispersion $\sigma_{\rm ISM} = 9\kms$ \citep{Braun09}. 

\citetalias{Reina-Campos17} showed that the predicted high-mass truncation of the ICMF in M31 has a large uncertainty due to the large variation in the gas conditions and the the star formation rate (SFR) over the last $300\Myr$. Specifically, \citet{Lewis15} show that the SFR in the star-forming ring varied by up to a factor of $\sim 4$ over the last $300\Myr$ with respect to the SFR during the most recent $25\Myr$. We include this effect in the uncertainty in the ICMF by correcting the gas surface density as follows. We assume that $\SigmaISM \propto \Sigma_{\rm SFR}$ and use the ratio of the peak SFR surface densities in each age bin considered to recover the peak gas surface density during the formation epoch of each cluster sample. The peak ISM surface density during a past epoch of cluster formation is then
\begin{equation}
    \SigmaISM(\tau_{\rm min}< t <\tau_{\rm max}) = \SigmaISM \times \frac{ \Sigma_{\rm SFR}(\tau_{\rm min}< t <\tau_{\rm max}) } { \Sigma_{\rm SFR}(0<t<25\Myr) } ,
\end{equation}
where $\tau_{\rm min}$ and $\tau_{\rm max}$ are the bracketing ages of the cluster sample, and the $\Sigma_{\rm SFR}$  values for $[\tau_{\rm min},\tau_{\rm max}] = [10,100]$ and $[100,300] \Myr$ are taken at a galactocentric radius of $12\kpc$ from Figure 6 of \citet{Lewis15}. The uncertainty in the gas surface density during the formation of the clusters of ages $\tau \in [\tau_{\rm min},\tau_{\rm max}]$ then lies in the range $[\SigmaISM(0<t<25\Myr), \SigmaISM(\tau_{\rm min}< t <\tau_{\rm max})]$.

\begin{figure*}
    \includegraphics[width=1.0\textwidth]{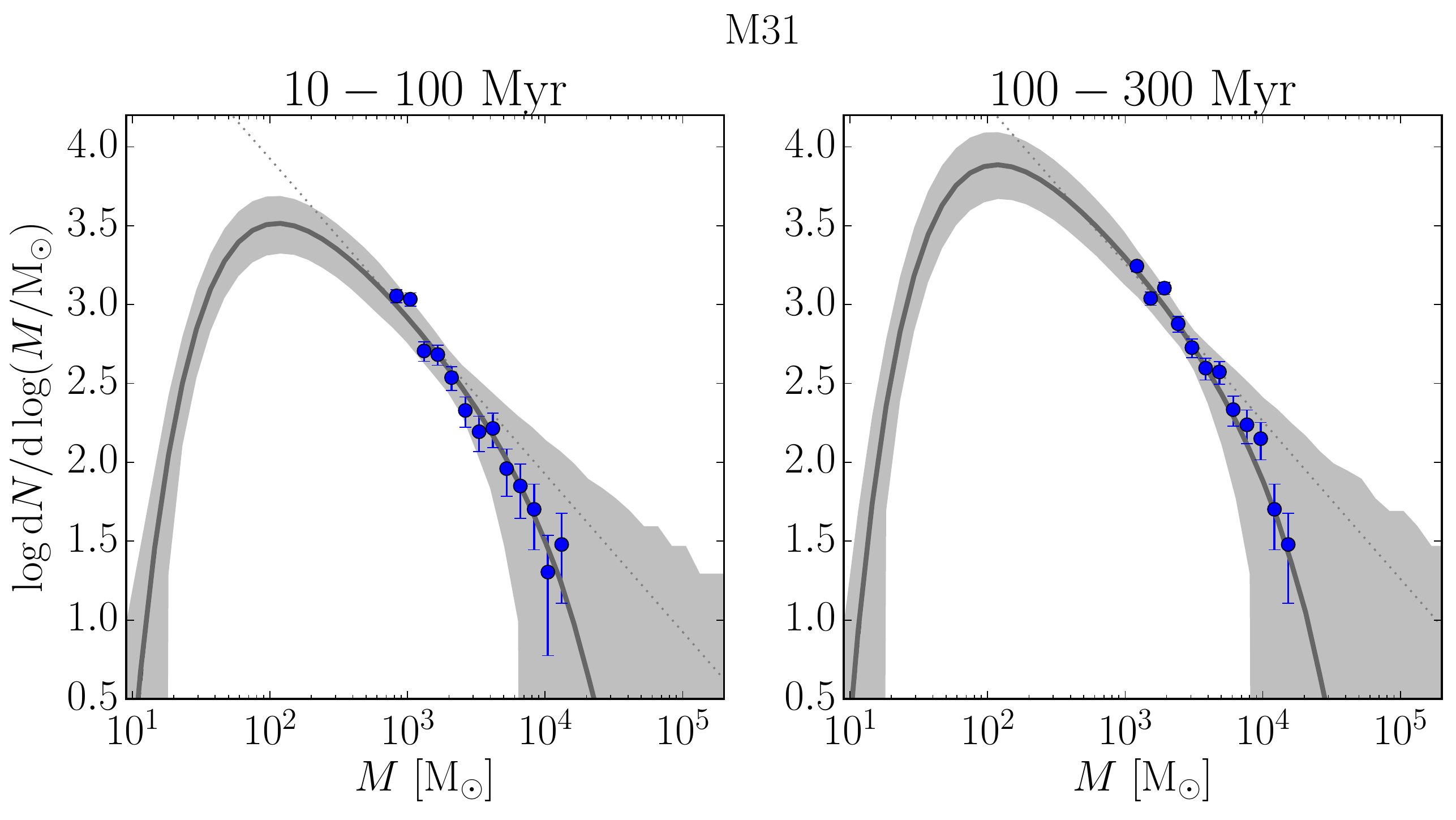}%
    \caption{Predicted and observed young CMF in M31 for clusters in two age bins (left: $\tau=10{-}100\Myr$; right: $\tau=100{-}300\Myr$). The prediction of our model, i.e.~equation~(\ref{eq:CMF}), is shown as a dashed line. The solid line and shaded area show the mean and the 2.5 to 97.5 percentile range of a Monte Carlo representation of the predicted ICMF. In this case the uncertainties account for variation in the ISM surface density of the star-forming ring in each of the formation epochs of the observed clusters. The dotted line shows a pure power law ICMF with $\beta = -2$. The points indicate the observations by \citet{Johnson17}. The model agrees well with the data down to the lowest masses probed by the observations.}
\label{fig:CMF_M31}
\end{figure*}

Figure~\ref{fig:CMF_M31} shows our predictions along with the observations by \citet{Johnson17}. The predicted minimum cluster mass is in the range $1.1{-}1.2\times10^2\Msun$ and the maximum cluster mass is  $4.2\times10^3{-}1.6\times10^6\Msun$ for clusters with ages up to $100 \Myr$. The minimum mass is in the range $1.1{-}1.2\times10^2\Msun$, while the maximum mass is $4.2\times10^3{-}2.1\times10^7\Msun$ for clusters with ages up to $300\Myr$. The spread of these mass scales comes exclusively from the variation of the conditions in the ISM inferred from the spatially-resolved star formation history.

The model is able to simultaneously fit the mass functions of clusters with ages $10< \tau <100\Myr$ and $100 < \tau < 300\Myr$ in M31. Unfortunately, the 50~per cent completeness limit of the \citet{Johnson17} data is well above the lower limit on the predicted minimum cluster mass of $\Mmin = 1.5\times10^2\Msun$. In spite of M31 having a lower gas ISM surface density and a lower SFR than the Milky Way, the dependence of the minimum mass on the ISM surface density is quite shallow in this region of the parameter space (see Figure~\ref{fig:paramspace}), resulting in a behaviour very similar to that of the solar neighbourhood. 

\subsection{The LMC}
\label{sec:lmc}

Because it is located only $\sim 50\kpc$ away, the LMC allows for the deepest determination of the ICMF beyond the Milky Way. With a star formation rate of $0.26\Msun~{\rm yr}^{-1}$ \citep{Kennicutt95}, which is about ten times lower than in the Milky Way (but much higher than the solar neighbourhood alone), the LMC presents a unique opportunity to test the ICMF model within a dwarf galaxy environment. This is also interesting, because the GC populations in nearby dwarfs are strikingly different from those of massive galaxies like the Milky Way (see Section~\ref{sec:intro}).

To apply our ICMF model to the LMC, we consider an ISM surface density of $\SigmaISM = 9.9\Msunpc2$ \citep{Staveley-Smith03}, a gas velocity dispersion of $\sigma_{\rm ISM} = 15.8\kms$ \citep{Kim98}, and the flat region of the rotation curve from \citet{Kim98} to derive an angular velocity of $\Omega \sim 0.031\Myr^{-1}$. For these parameters, the model predicts an ICMF with a low-mass truncation $\Mmin = 1.1\times10^2\Msun$ and a maximum cluster mass $\Mmax = 4.5\times10^4\Msun$. 

\begin{figure*}
    \includegraphics[width=1.0\textwidth]{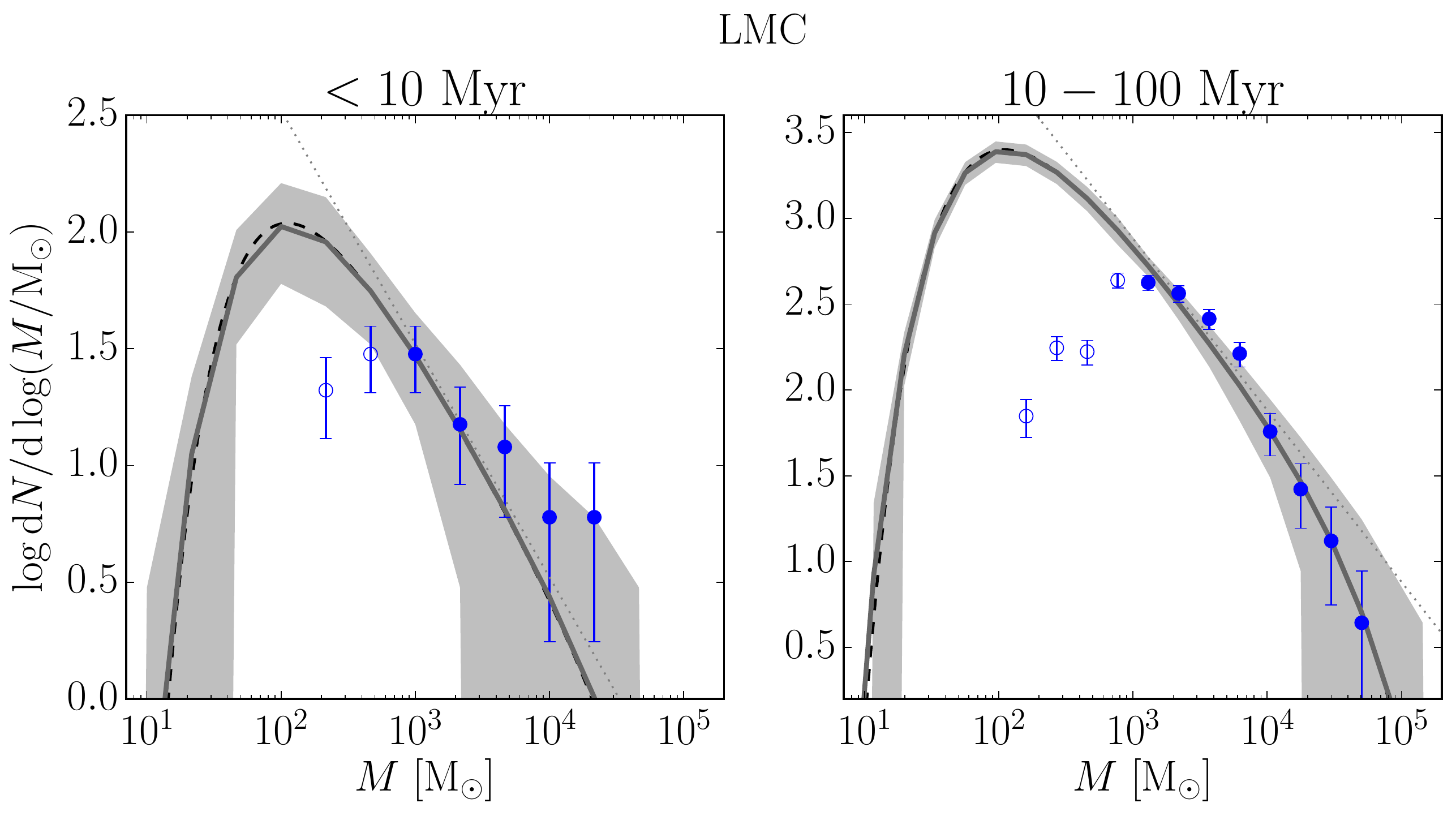}%
    \caption{Predicted and observed young CMF in the LMC in two age bins (left: $\tau < 10\Myr$; right: $\tau=10{-}100\Myr$). The prediction of the model, equation~(\ref{eq:CMF}) is shown as a dashed line. The solid line and shaded band show the mean and 2.5 to 97.5 percentile range of Monte Carlo samples drawn from the predicted ICMF, respectively. The points we obtained from the cluster catalogue by \citet{Popescu12}. Empty symbols denote clusters below the completeness limit set by the effect of fading. The observed young CMFs agree well with the predictions where the data are complete. }
\label{fig:CMF_LMC}
\end{figure*}

Figure~\ref{fig:CMF_LMC} shows the predicted ICMF and the observed young CMF derived from the \citet{Popescu12} catalogue of LMC clusters for clusters with ages $\tau < 10\Myr$ and $\tau < 100\Myr$. We use the mass completeness limit determined from the edge of the fading region in Figure 16 of \citet{Popescu12} for each age bin. The model is normalised to contain the same integrated number of clusters as the observations for each of the two age ranges. \citet{Popescu12} observe clusters in the LMC down to the lowest masses available for extragalactic objects, i.e.~$M\sim 10^{2.6}\Msun$ for cluster ages $<10\Myr$. However, this is still insufficiently deep to reach the turnover predicted by our model under the conditions of star formation in the LMC. Regardless, the ICMF of clusters with ages $< 100\Myr$ in our model shows very good agreement with the data above the completeness limit ($M \ga 10^3\Msun$ for this age range). 

\subsection{The Antennae galaxies}
\label{sec:antennae}

The Antennae galaxies are the nearest example of a pair of merging massive disc galaxies. They have been the subject of many studies due to their relative proximity of $\sim 20\Mpc$. Their interaction is driving a starburst with a star formation rate of $20\Msun~{\rm yr}^{-1}$ \citep{Zhang01} and resulting in the formation of very massive young clusters \citep{ZhangFall99, Whitmore10}. This is the ideal environment to study the effect of extreme ISM conditions during mergers on the ICMF.

To derive the predicted ICMF, we use the average observed gas ISM surface density and velocity dispersion across the discs, $\SigmaISM \sim 200\Msun\pc^{-2}$ and $\sigma_{\rm ISM} \sim 30\kms$ \citep{Zhang01}. For the angular velocity we assume it to be $\Omega\sim 0.07\Myr^{-1}$, based on a rough estimate of the gas rotation velocity gradient ($\sim 67\kms\kpc^{-1}$ in a linearly rising rotation curve) from the \HI velocity field in \citet{Hibbard01}. Using these values, the model predicts low- and high-mass truncation masses of $\Mmin = 2.2\times10^2\Msun$ and $\Mmax = 2.5\times10^7\Msun$.

Naturally, our estimate of $Q$ and gas volume density in this merging system may be inaccurate as it is based on the assumption of a rotating disc in hydrostatic equilibrium. However, \citet{Meng18} use simulations to suggest that the Toomre criterion is still applicable to irregular high redshift galaxies, which do undergo frequent mergers. This implies that our estimate of $Q$ for the Antennae might be reasonable. Moreover, as shown in Figure~\ref{fig:paramspace}, the minimum mass is driven primarily by the ISM surface density, with a dependence on Toomre $Q$ only for very high gas surface densities $\SigmaISM > 3\times10^3\Msunpc2$ (see Section~\ref{sec:environments}). 
\begin{figure*}
    \includegraphics[width=1.0\textwidth]{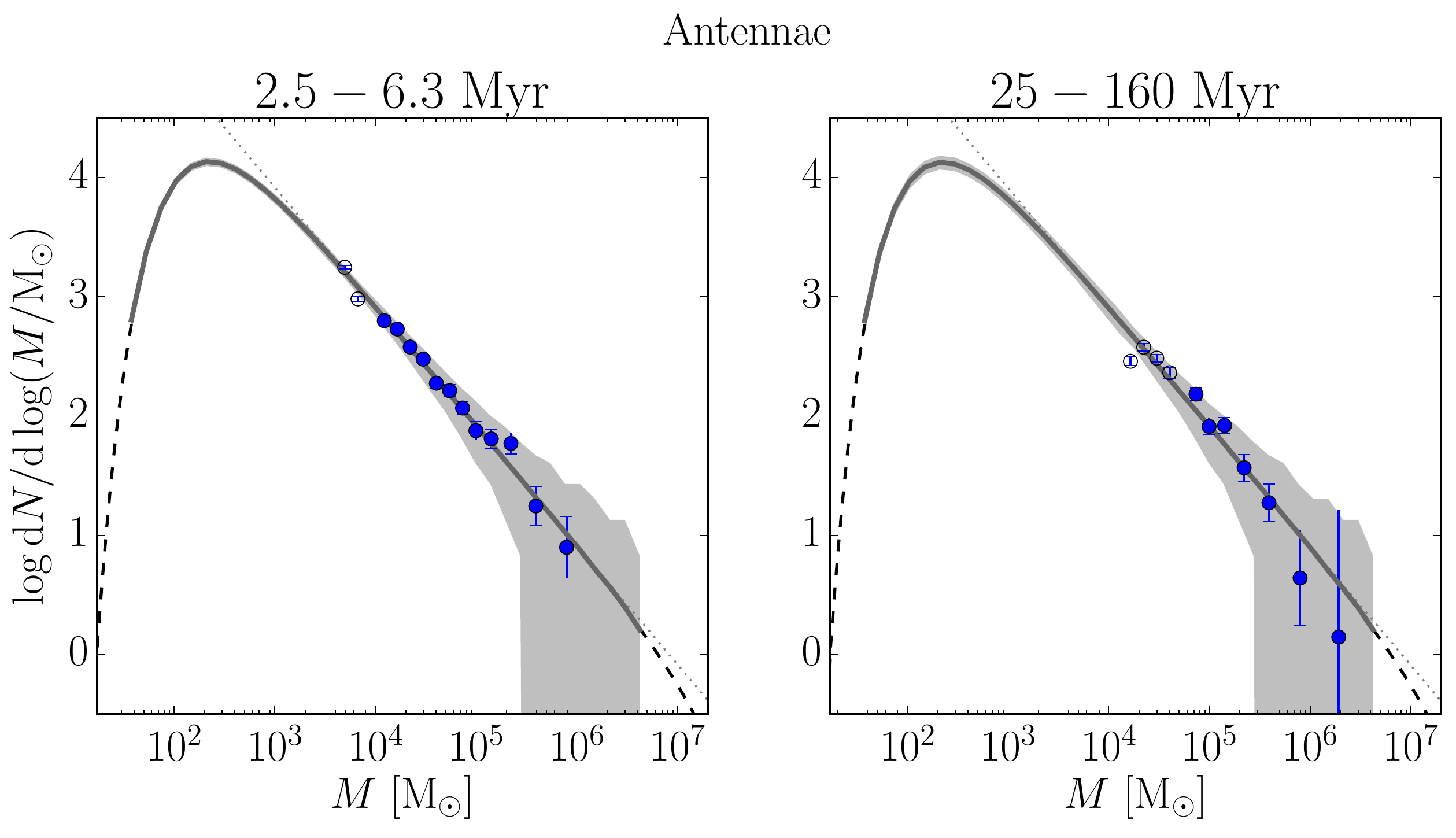}%
    \caption{Predicted and observed young CMF in the Antennae galaxies for clusters in two age ranges (left: $\tau=2.5{-}6.3\Myr$; right: $\tau=25{-}160\Myr$). The prediction of our model, equation~(\ref{eq:CMF}) is shown as a dashed line. The solid line and shaded band show the mean and 2.5 to 97.5 percentile range of a Monte Carlo representation of the predicted ICMF. The points are the observations from \citet{ZhangFall99}, with open symbols denoting bins which are affected by incompleteness or stellar contamination. The model agrees well with the observations above the completeness limit. }
\label{fig:CMF_Ant}
\end{figure*}

The observed young CMF in the Antennae is determined by \citet{ZhangFall99} using HST observations. Figure~\ref{fig:CMF_Ant} shows the comparison of our predicted ICMF with observations. To normalise the model ICMF, we use the total completeness-corrected number of clusters inferred from the \citet{ZhangFall99} data for each of the two age intervals, $2.5 < \tau < 6.3\Myr$ and $25 < \tau < 160\Myr$. The observed CMFs are complete down to $7.9\times10^3\Msun$ and $2.5\times10^4\Msun$ for the respective age intervals. The distribution of Monte Carlo samples obtained from the model agrees very well with the observations down to the completeness limit, which is well above the predicted minimum bound cluster mass.   

The young CMF in the Antennae is perfectly fit by a power-law in the range $\sim 10^4{-}10^6\Msun$. This range is well above the minimum mass predicted by our model, $\Mmin = 2.2\times10^2\Msun$, and consistent with the statistically observable maximum mass given the size of the cluster sample. The predictions for the minimum cluster mass are thus consistent with the limits provided by observations in the regime of local galaxies with high gas and star formation surface densities.

\subsection{The CMZ of the Milky Way and the M82 nuclear starburst}
\label{sec:cmzs}

Circumnuclear (starbursting) rings are another example of extreme star formation environments commonly found in nearby massive galaxies. They are characterised by a narrow ring of dense molecular gas located near the centre of the galaxy. These are ideal environments to test very high galactic shear ($\Omega \ga 1\Myr^{-1}$) and ISM surface density ($\SigmaISM \ga 5\times 10^2\Msunpc2$) conditions where our model predicts narrow CMFs (see Figure~\ref{fig:paramspace}). 

The CMZ is the region located within the central $\sim 500\pc$ of the Milky Way. It has a surprisingly high molecular gas surface density given its low SFR $\sim 0.09\Msun\yr^{-1}$ \citep{Longmore13,Barnes17}. Its high gas surface density ($\sim 100$ times higher than in the solar neighbourhood), as well as its location in a region dominated by shearing motions makes it ideal for studying the star-forming conditions in an environment similar to that of high-redshift galaxies \citep{Kruijssen13}. To predict the ICMF in the CMZ, we consider an ISM surface density $\SigmaISM \sim 10^3\Msun\pc^{-2}$ and a gas velocity dispersion of $\sigma_{\rm ISM} = 5\kms$ \citep{Henshaw16}, and use the enclosed mass profile from \citet{Kruijssen15a} to obtain the angular velocity and the Toomre $Q$ parameter at a radius of $60\pc$ \citep[the innermost radius of the molecular stream, cf.][]{Molinari11,Kruijssen15a}, resulting in $\Omega = 2.04\Myr^{-1}$ and $Q = 1.32$.  

\begin{figure*}
    \includegraphics[width=1.0\textwidth]{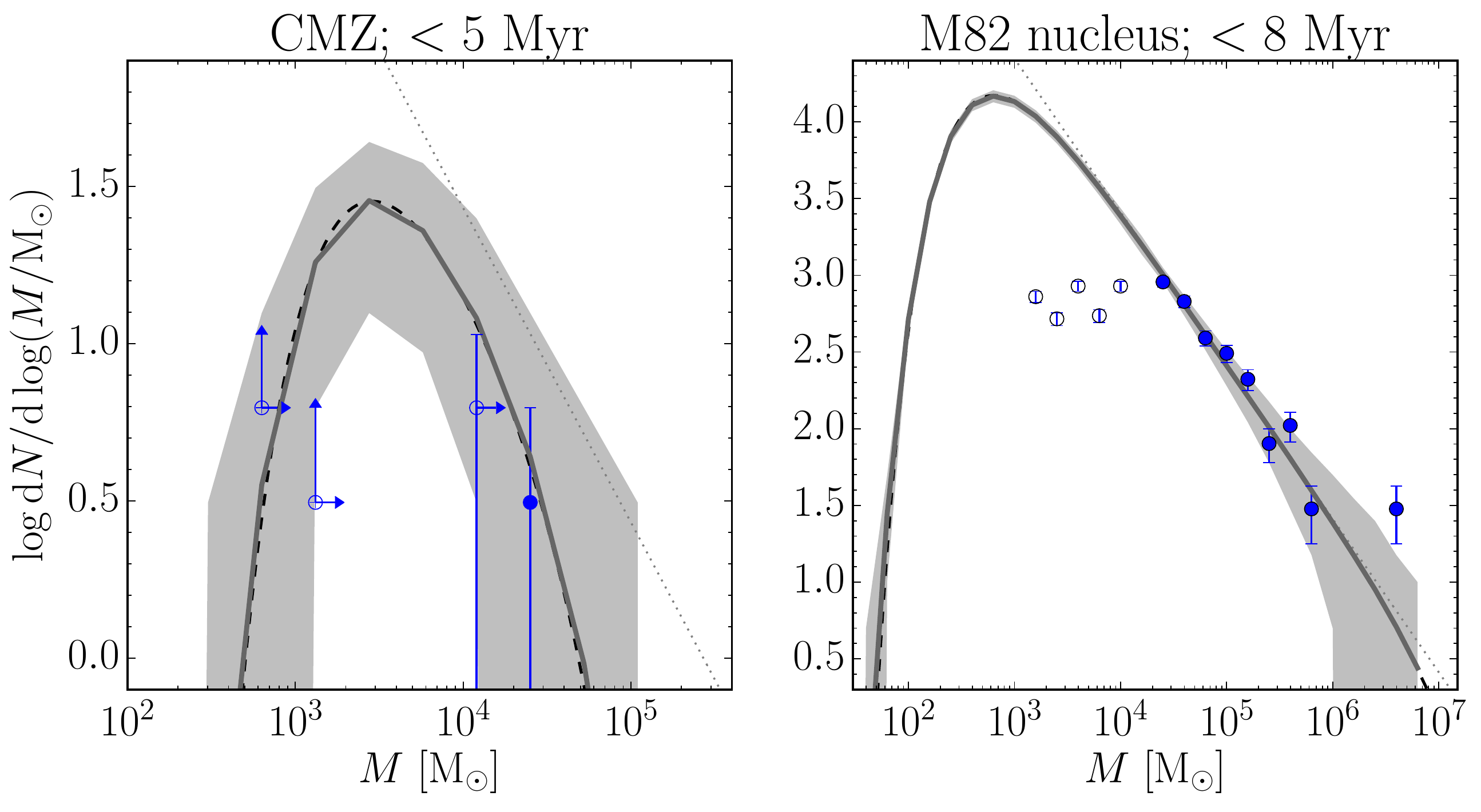}%
    \caption{Predicted and observed young CMF in the CMZ of the Milky Way and in the nearby nuclear starburst in M82. {\it Left}: the prediction of model, equation~(\ref{eq:CMF}), for clusters with ages $\tau < 5\Myr$ in the CMZ is shown as a dashed line. The solid line shows the result of Monte Carlo sampling of the predicted ICMF, and the points represent the only two unembedded clusters observed in the CMZ \citep{PortegiesZwart10}, together with lower limits based on the embedded proto-clusters \citep{GinsburgKruijssen18}. Open symbols represent lower limits in the horizontal or vertical direction, with vertical arrows indicating bins affected by incompleteness, and horizontal arrows showing bins with lower mass limits from proto-clusters. The extremely narrow predicted ICMF, which in our model is due to high ISM surface densities and strong centrifugal support, is consistent with the absence of low- and high-mass clusters. {\it Right}: Comparison between the predicted young CMF in the nucleus of M82 with observations for clusters with ages $\tau \leq 8 \Myr$. Empty symbols denote data below the completeness limit. }
\label{fig:CMF_CMZ}
\end{figure*}

The left panel of Figure~\ref{fig:CMF_CMZ} shows the model prediction for the mass function in the CMZ. In this case, because of its location near the galactic centre, the \citetalias{Reina-Campos17} model predicts that the maximum cloud mass is limited by the mass enclosed in the region unstable to centrifugal forces, with $\Mmax = 3.0\times10^4\Msun$. In addition, very high gas surface densities allow more massive clouds to collapse into single bound clusters, with a minimum cluster mass $\Mmin = 3.2\times10^3\Msun$.  This combination of high ISM surface density and strong shear thus results in a very narrow ICMF, with a width of less than one decade in mass.  

Due to high extinction towards the galactic centre, it is difficult to obtain the young CMF in the CMZ. To compare with the model, we use the masses of the only young clusters found in the region (the Arches and Quintuplet) from \citet{PortegiesZwart10}, and include as lower limits the mass estimates for the five embedded ``proto-clusters'' from Table 1 of \citet{GinsburgKruijssen18}. To normalise the predicted ICMF we use the SFR determination by \citet{Barnes17}, who find that several methods agree to within a factor of 2 with a mean value of the total SFR $= 0.09\Msunyr$. Multiplying this by the observed CFE in Sgr B2, $ 37 \pm 7$ per cent \citep{GinsburgKruijssen18}, and by the width of the cluster age interval, $\sim 5\Myr$, the total mass of stars in young clusters is $\approx 1.66\times10^5\Msun$.  Figure~\ref{fig:CMF_CMZ} shows that the CMZ is one of the star formation environments in the Local Universe in which the ICMF is expected to deviate most strongly from the traditionally-assumed Schechter form with a lower limit at $10^2\Msun$. Despite the poor statistics in this region, our prediction is consistent with the masses of observed clusters and with the lower limits set by proto-clusters. 

As an example of a starburst environment in an external galaxy with a well-sampled mass function, we show in the right panel of Figure~\ref{fig:CMF_CMZ} the observed young CMF in the central starburst region of M82, along with the predictions of our model. To obtain the predictions in the nuclear region, we use the median inferred molecular gas column density $\SigmaISM = 500\Msunpc2$ from \citet{Kamenetzky12}, an angular velocity of $\Omega = 0.23\Myr^{-1}$, and a gas velocity dispersion of $\sigma_{\rm ISM} = 60\kms$. The angular velocity is calculated for the $\sim 450\pc$ central region using the M82 mass model from \citet{Martini18}. The ISM velocity dispersion was calculated in the same region using the map of CO velocity dispersion in Fig. 5 of \citet{Leroy15}. We show the CMF obtained by \citet{Mayya08} for clusters in the nuclear region with ages $\tau \leq 8\Myr$. 

The model predicts $\Mmin = 6.4\times10^2\Msun$ and $\Mmax = 1.5\times10^7\Msun$. The CMF prediction reproduces the observed mass function of young clusters in M82 down to the completeness limit of the observations, $M\sim 2\times10^4\Msun$, which lies above the theoretical minimum cluster mass.  

\subsection{The cluster formation efficiency}
\label{sec:cfe}

The fraction of stars that form in bound clusters relative to the field is key for understanding the formation of cluster populations and for their use as tracers of galaxy evolution \citep[e.g.][]{Bastian08, Adamo11, Kruijssen12b, Cook12, Hollyhead16, Johnson16, Messa18}. To determine the cluster formation efficiency (CFE) observationally, the ICMF must be integrated down to its low-mass truncation, which is traditionally assumed to be $M\sim 10^2\Msun$ \citep{LadaLada03, Lamers05}. To evaluate the effect of the environmental variation of the minimum cluster mass on the measurement of the cluster formation efficiency, we now compare the observationally determined CFE using the minimum mass model, equation~(\ref{eq:boundcondISM}), with the CFE obtained assuming the traditional $10^2\Msun$ truncation. 

Table~\ref{tab:CFE} summarises the values of the minimum and maximum cluster masses obtained using our model. To highlight the relative change in the CFE estimates that results from using the environmentally-dependent minimum cluster mass (equation~\ref{eq:boundcondISM}) instead of the traditional $10^2\Msun$ value, we show the ratio of the two estimates. Following \citet{Bastian08}, we use the following definition of the CFE for a cluster sample with an upper age limit $\tau$: 
\begin{equation}
    \Gamma = \frac{ \int_{0}^{\infty} {\rm ICMF}(M, \Mmin, \Mmax) M~{\rm d}M }{ \tau \times {\rm SFR}(<\tau) } ,
    \label{eq:Gamma}
\end{equation}
where the ICMF is obtained using equation~(\ref{eq:CMF}) with the value of $\Mmin$ and $\Mmax$ taken from Table~\ref{tab:CFE}. To calculate the CFE using the traditional $10^2\Msun$ truncation we evaluate equation~(\ref{eq:Gamma}) with $\Mmin=10^2\Msun$. Note that the denominator in equation~(\ref{eq:Gamma}) drops out when taking the ratio of the two CFE values, $\Gamma(\Mmin)/\Gamma(10^2\Msun)$.

\begin{table*}
	\centering
	\caption{Comparison of the observed cluster formation efficiency determined using the minimum cluster mass model versus assuming the traditional $10^2\Msun$ low mass truncation of the ICMF.}
	\label{tab:CFE}
	\begin{tabular}{lccc} 
		\hline \hline
         Environment         & $\Mmin~[\Msunns]$ & $\Mmax~[\Msunns]$  & $\Gamma({\Mmin})/\Gamma({10^2\Msun})$ \\
		\hline
		Solar neighbourhood  & $1.1\times10^2$ & $2.8\times10^4$ & 0.97 \\ 
		M31                  & $1.1\times10^2$ & $8.2\times10^4$ & 0.98 \\ 
		LMC                  & $1.1\times10^2$ & $4.5\times10^4$ & 0.98 \\ 
		Antennae             & $2.2\times10^2$ & $2.5\times10^7$ & 0.93 \\ 
		CMZ                  & $3.2\times10^3$ & $3.0\times10^4$ & 0.31 \\ 
		M82                  & $6.4\times10^2$ & $1.5\times10^7$ & 0.86 \\ 
		\hline \hline
    \end{tabular}
    \begin{tabular}{l}
		Notes: the ``traditional'' ICMF assumes a Schechter function with a lower mass limit at $10^2\Msun$. 
	\end{tabular}
\end{table*}

The derived cluster formation efficiencies obtained with the environmentally dependent minimum mass model are $\sim 2-70$ per cent lower in the representative set of environments shown in Table~\ref{tab:CFE}. Taking the nucleus of M82 as a representative starburst with $\Mmin \gg 10^2\Msun$, the predicted CFE is $\sim 15$ per cent lower than using the traditional ICMF. In the CMZ the CFE is overestimated by up to $\sim 70$ per cent when assuming a traditional ICMF with $\Mmin=100\Msun$. The effect of the environmental variation of the minimum cluster mass should thus be taken into account when comparing observations with theoretical predictions of the CFE.

\section{Implications for GC formation}
\label{sec:GCs} 

The ISM conditions in the progenitors of present-day galaxies are typically extremely difficult to study in detail at high redshift. Even when this is possible, the star-forming conditions in the progenitor can only be matched to their local counterparts in a statistical sense. As we have shown, the initial mass distribution of star clusters contains an imprint of their birth environment in the form of the variation of the minimum and maximum masses. It is possible that the present-day CMF of those clusters that survive for many billions of years could preserve a record of these conditions.

The mass distribution of present-day GC populations offers a unique window into the physical environments of their host galaxies in the early Universe. There is now growing evidence that GCs can be understood as products of regular star formation in the high pressure conditions of high-redshift galaxies, shaped by billions of years of dynamical evolution within their host galaxy \citep{KravtsovGnedin05, Elmegreen10, Kruijssen15b, Lamers17, Forbes18, Pfeffer18, Kruijssen19a}. 

If the ICMF can be reconstructed from the observed globular cluster mass function (GCMF), then our model can be inverted to constrain the star-forming environments (i.e.~gas surface density and angular velocity) that gave rise to such a mass function. The GCMF is therefore an ideal tool to probe the otherwise inaccessible early star-forming conditions in present-day galaxies.  

\subsection{The progenitors of dwarf galaxies -- the Fornax dSph}

The progenitors of dwarf galaxies are observationally inaccessible at high redshift, and the only information that may be used to constrain their formation and evolution comes from resolved star formation histories in the Local Group \citep[e.g.][]{Weisz14a, Weisz14b}. These provide information on the dwarfs' ancient stellar populations, but do not constrain the star formation environments (gas surface density and angular velocity) that would be imprinted in the CMF as predicted by our model. Determining the ICMF may therefore provide additional constraints on the early star-forming conditions in galaxies. 

Several nearby dwarf galaxies are extremely efficient at forming GCs, especially at early cosmic epochs, compared to more massive galaxies like the Milky Way. \citet{Larsen12} found that $\leq 20$ per cent of the stars with metallicity $[\rm{Fe/H}]<-2.0$ reside in Fornax's GCs. Similarly large fractions are also observed in other dwarfs including IKN and WLM \citep{Larsen14}. These observations are very difficult to explain using a traditional Schechter ICMF with a minimum cluster mass of $10^2\Msun$, because the amount of mass lost from the surviving GCs can account for up to half of the total mass of low-metallicity stars in the entire galaxy, leaving little room for stars from the remnants of the numerous low-mass clusters that did not survive. Clearly, a narrow ICMF with a high minimum mass could explain this puzzling observation.

To evaluate whether this scenario is feasible physically, we invert our model for the ICMF to find the star-forming conditions that led to the formation of the Fornax GCs $\sim 10-12$ Gyr ago \citep{deBoerFraser16}. This requires that we make an assumption about how well the current GC population represents the initial cluster mass distribution. Since the metal-poor Fornax GCs are all massive ($M \ga 10^5\Msun$), the mass loss due to evaporation and tidal shocks are expected to be sub-dominant (or similar) relative to mass loss due to stellar evolution \citep{Reina-Campos18}. Using the most conservative assumption of mass loss due to only stellar evolution, two extreme scenarios will bracket the range of possible ICMFs: 
\begin{enumerate}
    \item \emph{Minimum CFE} model: none of the present field stars were originally born in clusters and the observed GCs represent the initial CMF (i.e.~the CFE is the ratio of mass in GCs to mass in low-metallicity stars, $\sim 20$ per cent).
    \item \emph{Maximum CFE} model: all the low-metallicity field stars in the galaxy originated from disrupted bound clusters (i.e.~the CFE is 100 per cent).
\end{enumerate}

We may then fit the inferred ICMFs from each model with equation~(\ref{eq:CMF}) to recover the minimum and maximum cluster masses. The minimum mass can be used to invert equation~(\ref{eq:boundcondISM}) and solve for the star-forming conditions (i.e.~the ISM surface density and angular rotation velocity) of Fornax at the epoch of formation of its GC population. An additional constraint on the gas surface density and the angular velocity can be obtained from inverting equation~(10) in \citetalias{Reina-Campos17} (modified to account for IMF sampling following Appendix~\ref{sec:appendix}) for the maximum cluster mass. Together, these two independent constraints should significantly narrow down the range of physical conditions that produced the metal-poor Fornax GCs.

\begin{figure}
    \includegraphics[width=1.0\columnwidth]{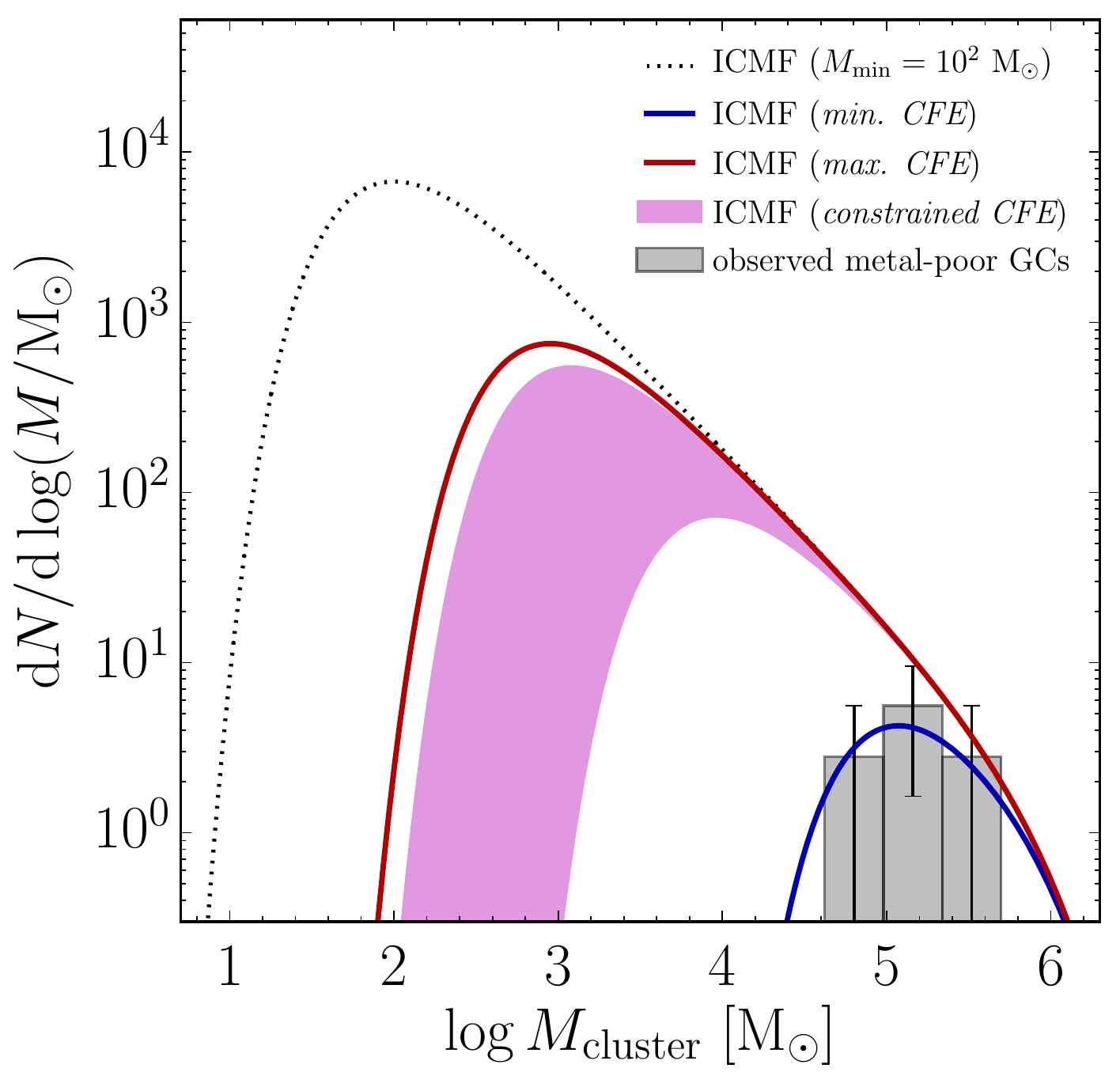}
    \caption{Prediction for the ICMF in the Fornax dSph at the time of formation of its metal-poor GC population $10-12\Gyr$ ago. The histogram with error bars shows the observed GCMF of the four GCs with $[\rm{Fe/H}]<-2$ (corrected for mass loss due to stellar evolution, see \citealt{deBoerFraser16}). The blue curve is the ICMF in the \emph{minimum CFE} model, where the only clusters formed are the presently observed GCs (i.e.~the CFE is $M_{\rm tot}^{\rm GCs}/M_{\rm Fornax}$). The red curve is the ICMF for the \emph{maximum CFE} model, where \emph{all} the low-metallicity stars in the galaxy came from disrupted low-mass clusters (i.e.~the CFE is 100 per cent). The purple shaded region represents the range of models constrained by the CFE predicted by \citet{Kruijssen12b} for $Q$ in the range $0.5{-}3$. For reference, the dotted line shows the traditional ICMF with $\Mmin=10^2\Msun$.}
\label{fig:CMF_Fornax}
\end{figure}
Figure~\ref{fig:CMF_Fornax} shows the observed GCMF and the resulting ICMF for each of the two bracketing scenarios. Here we used the \emph{birth} GC masses derived by \citet{deBoerFraser16} using color-magnitude diagrams and assuming a \citet{Kroupa01} IMF. For the \emph{minimum CFE} model we simply fit our model ICMF (equation~\ref{eq:CMF}) to the observed Fornax GCMF using the Maximum Likelihood Estimator \citep{Fisher1912} and equation~(\ref{eq:CMF}) with uniform priors on $\Mmin$ and $\Mmax$ in the region defined by the condition
\begin{equation}
    M_{\rm GCs}^{\rm lower} \leq \int_0^{\infty} \mathrm{ICMF}(\Mmin,\Mmax,M) M {\mathrm d}M \leq M_{\rm GCs}^{\rm upper} ,
\end{equation} 
where $M_{\rm GCs}^{\rm lower}$ and $M_{\rm GCs}^{\rm upper}$ are the lower and upper estimates of the total mass of the four low-metallicity GCs from \citet{deBoerFraser16} respectively. This condition merely states that the total mass under the ICMF should agree with the observed total mass of the GCs within the limits set by the observational errors. The best-fit model is shown in Figure~\ref{fig:CMF_Fornax} and its parameters are $\Mmin = 1.4^{+1.4}_{-0.8}\times10^5\Msun$ and $\Mmax = 8.1^{+109.6}_{-5.6}\times10^5\Msun$, where the errors correspond to the parameter values for which the likelihood drops by a factor of $1/e$. These numbers show that the number of GCs in Fornax is too small to put meaningful constraints on any possible high-mass truncation of the ICMF.

Next, to obtain the minimum cluster mass for the \emph{maximum CFE} model, we decrease $\Mmin$ in the \emph{minimum CFE} model -- while holding $\Mmax$ fixed -- until the total mass under the ICMF equals the total mass of Fornax at $[{\rm Fe/H}] < -2$, which is estimated to be $\sim 5$ times the total GC mass \citep{Larsen12, deBoerFraser16}. Figure~\ref{fig:CMF_Fornax} illustrates the budget problem of the traditional ICMF: to avoid overproducing the stellar mass of the entire galaxy, the minimum cluster mass should be larger than $\sim 900\Msun$, assuming a well-sampled ICMF.

\begin{figure*}
    \includegraphics[width=0.95\textwidth]{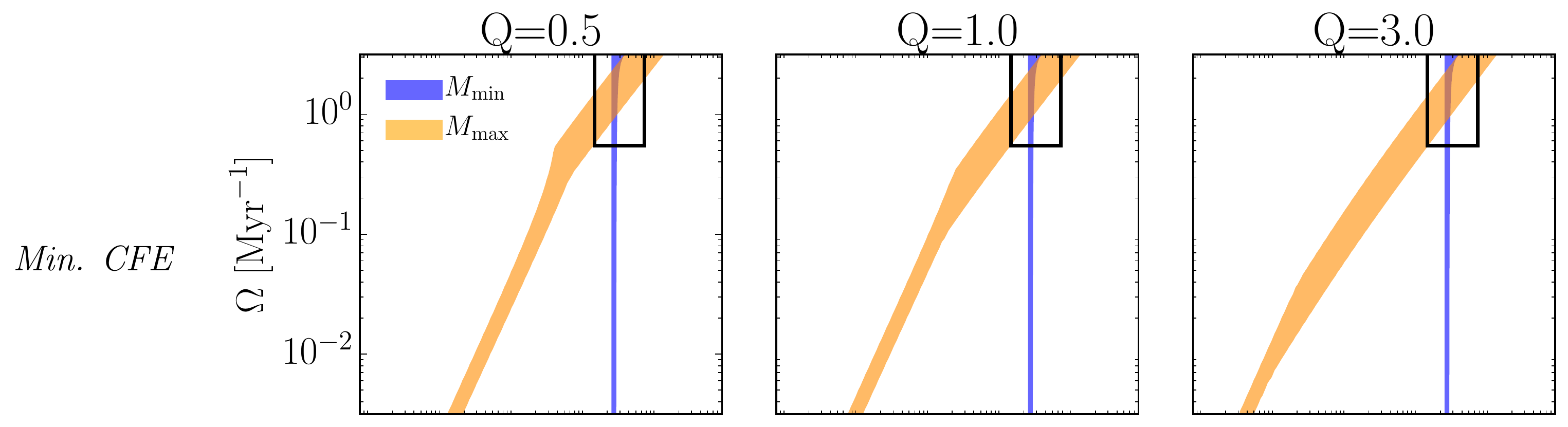}
    \includegraphics[width=0.95\textwidth]{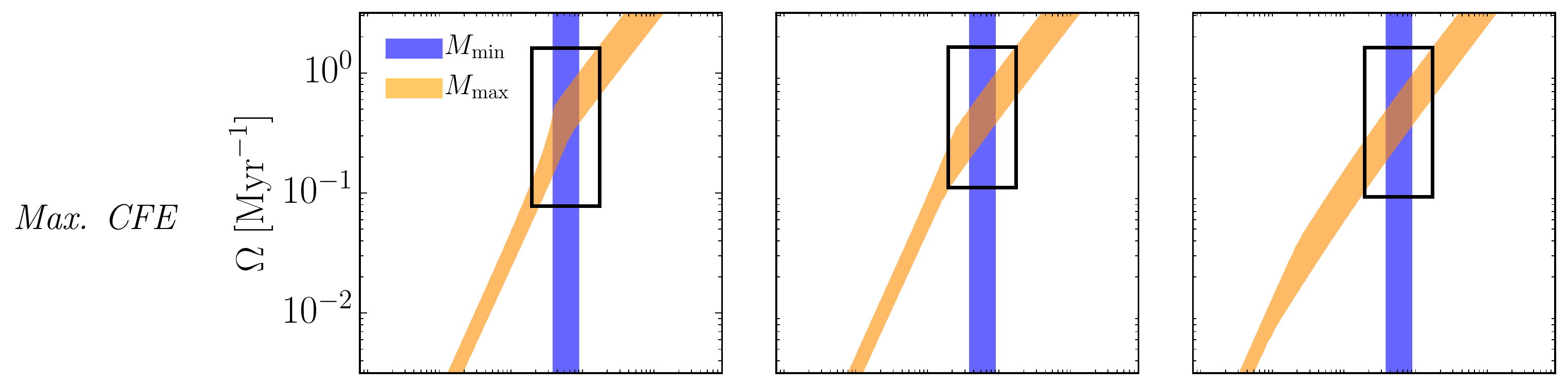}
    \hspace*{+0.1cm}\includegraphics[width=0.95\textwidth]{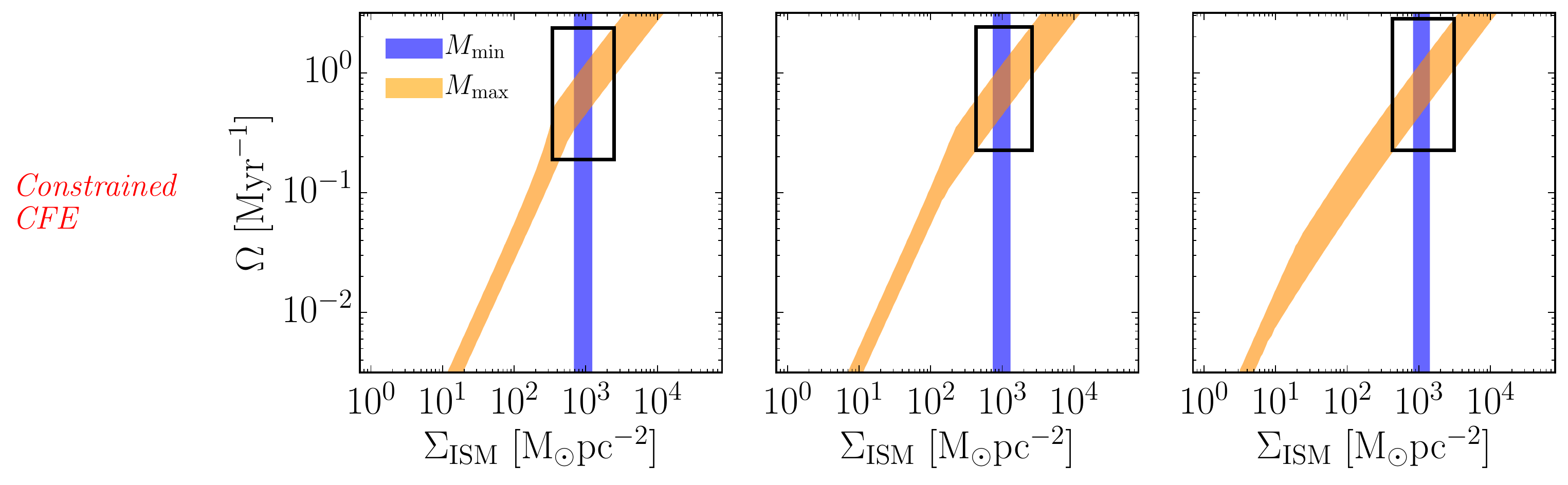}
\caption{Star-formation environment in the Fornax dSph during the epoch of formation of its GCs, $\sim 10-12~\Gyr$ ago. The parameter regions in the ISM surface density and angular velocity plane that give rise to the ICMFs shown in Figure~\ref{fig:CMF_Fornax} are indicated in blue for $\Mmin$ and orange for $\Mmax$, for a representative range of values of the Toomre $Q$ parameter, $Q \in [0.5,1.0,3.0]$. \emph{Top row:} \emph{minimum CFE} model where all the clusters formed survived to the present day. \emph{Middle row:} \emph{maximum CFE} model where all the low-metallicity stars in the galaxy were born within clusters (i.e.~the CFE is 100 per cent). \emph{Bottom row:} \emph{constrained CFE} model where the minimum cluster mass is determined by matching the CFE predicted by \citet{Kruijssen12b} as a function of ISM surface density, angular velocity and $Q$ parameter. Only the regions where both the contours of the minimum and maximum masses overlap (as indicated by boxes), represent physical solutions. This figure shows that the extremely high efficiency of GC formation in Fornax could be explained by very high ISM surface densities and shear $\sim 10-12~\Gyr$ ago. }
\label{fig:Fornax_conditions}
\end{figure*}

Is it now possible to recover, using the two ICMFs in Figure~\ref{fig:CMF_Fornax}, the location in the parameter space of ISM surface density, angular velocity and Toomre $Q$ that produced the observed GCMF. This reconstruction of the ISM surface density and angular velocity is shown in the top and middle rows of Figure~\ref{fig:Fornax_conditions} for a range of values $Q \in [0.5,1.0,3.0]$. Since the minimum and maximum masses have different environmental dependences (blue and orange shaded regions), they produce independent constraints on the environmental conditions for a fixed value of $Q$. The region where they overlap corresponds to the only physical solution for ISM surface density and angular velocity.

Interestingly, Figure~\ref{fig:Fornax_conditions} shows that only a very narrow range of conditions at fixed Toomre $Q$ yield a physical solution (where the minimum and maximum mass solution regions overlap). The largest uncertainty remaining in the predictions comes from the uncertainty in the CFE. 

To obtain tighter, self-consistent constraints on the ICMF, we further include the \citet{Kruijssen12b} model for the CFE. This model predicts, given the ISM surface density, angular velocity, and $Q$, the fraction of stars that form in bound clusters relative to the total amount of star formation. We can adjust the minimum mass in our model ICMF until the total mass in clusters matches the \citet{Kruijssen12b} CFE prediction obtained from the ISM conditions corresponding to the chosen value of $\Mmin$ and the best-fitting value of $\Mmax$. In other words, we are solving the equation
\begin{equation}
  \Gamma^{\rm K12}(\SigmaISM,\Omega,Q) = \frac{ \int_{0}^{\infty} {\rm ICMF}(M_{\rm min},M_{\rm max},M) M {\rm d}M }{ M_{\rm Fornax} } 
\end{equation}
together with equation~(\ref{eq:boundcondISM}) and equation~(26) of \citet{Kruijssen12a} for the CFE\footnote{Note that in this paper we define the CFE as the bound fraction of star formation. This is not the same notation used by \citet{Kruijssen12a}, where the CFE also includes the effect of early cluster disruption by tidal shocks.} implicitly for $\Mmin$, $\SigmaISM$, and $\Omega$, for a fixed assumed value of $Q$. Here, the ICMF is given by equation~(\ref{eq:CMF}) with the maximum mass held constant at the value $\Mmax = 8.1\times10^5\Msun$ obtained from the MLE fit to the GCMF, and $M_{\rm Fornax} = 4.49\times10^6\Msun$ is the total mass in low-metallicity stars in the galaxy \citep{deBoerFraser16}. The solution for $Q$ values in the range $0.5 \leq Q \leq 3.0$ is given by a CFE between 71~and 80~per cent for minimum masses $\Mmin = 1.2 - 9.4\times 10^3\Msun$ including $1\sigma$ uncertainties.

The range of solutions for the self-consistent, \emph{constrained CFE} model (including uncertainties) is indicated with a purple shaded region in Figure~\ref{fig:CMF_Fornax}, and the recovered conditions are shown on the bottom row of Figure~\ref{fig:Fornax_conditions}. Assuming the typical $Q \sim 0.5$ conditions of high-redshift galaxies, these results indicate that $\sim 10-12~\Gyr$ ago the progenitor of Fornax was forming stars in a high-surface density ISM (with $\SigmaISM \simeq 700{-}1100\Msunpc2$) and strong shearing motions ($\Omega \simeq 0.4{-}1.1\Myr^{-1}$). These conditions caused an increase in the minimum cluster mass (compared to nearby disc galaxies) to $\Mmin = 1.2{-}5.8\times10^3\Msun$, and a high cluster formation efficiency of $\Gamma \sim 80$ per cent.

These recovered conditions are quite typical of local (nuclear) starburst galaxies, as indicated by the white circles in Figure~\ref{fig:paramspace}. This implies that a galactic environment similar to observed present-day nuclear starbursts could explain the large number of low-metallicity stars that belong to the GC systems of dwarf galaxies like Fornax, IKN, and WLM. Previously, \citet{Kruijssen15b} explained the high number of GCs in Fornax by an early galaxy merger that cut short the initial phase of rapid GC disruption due to tidal interactions with molecular clouds in the natal disc, instead redistributing the GCs into the gas-poor spheroid. With our model, an early galaxy merger is no longer required to explain the extremely high specific frequency at low metallicities in Fornax, even if it remains a possibility.

An interesting implication of this result is that Fornax (and due to the similarities in GC populations, also IKN and WLM) may all have undergone significant subsequent expansion of their stellar components, presumably due to the change of gravitational potential following the blow-out of the residual gas by stellar feedback. We plan to investigate the physics driving this expansion further in a follow-up paper.

\subsection{High-redshift star-forming galaxies}

Because of their location in the high ISM surface density ($\SigmaISM \ga 10^2\Msunpc2$) region of the parameter space in Figure~\ref{fig:paramspace}, high-redshift star-forming galaxies observed at $z=2-3$ are predicted by our model to have ICMFs with minimum masses that are factors of several to orders of magnitude larger that in local spirals. Furthermore, the \citetalias{Reina-Campos17} maximum cluster mass model predicts that the largest gas clumps will have masses larger than $10^9\Msun$, with clusters as massive as $10^8\Msun$ that will quickly spiral into the center through dynamical friction. 

To make a prediction for the ICMF of this class of galaxies, we select zC406690 ($z = 2.196$) from the catalogue provided by \citet{Tacconi13}. This object was chosen by \citetalias{Reina-Campos17} because it represents the average properties of high-redshift star-forming galaxies, its kinematics are dominated by rotation, and both its global properties and its molecular clump masses have been measured \citep{genzeletal11}.

\begin{figure}
    \includegraphics[width=1.0\columnwidth]{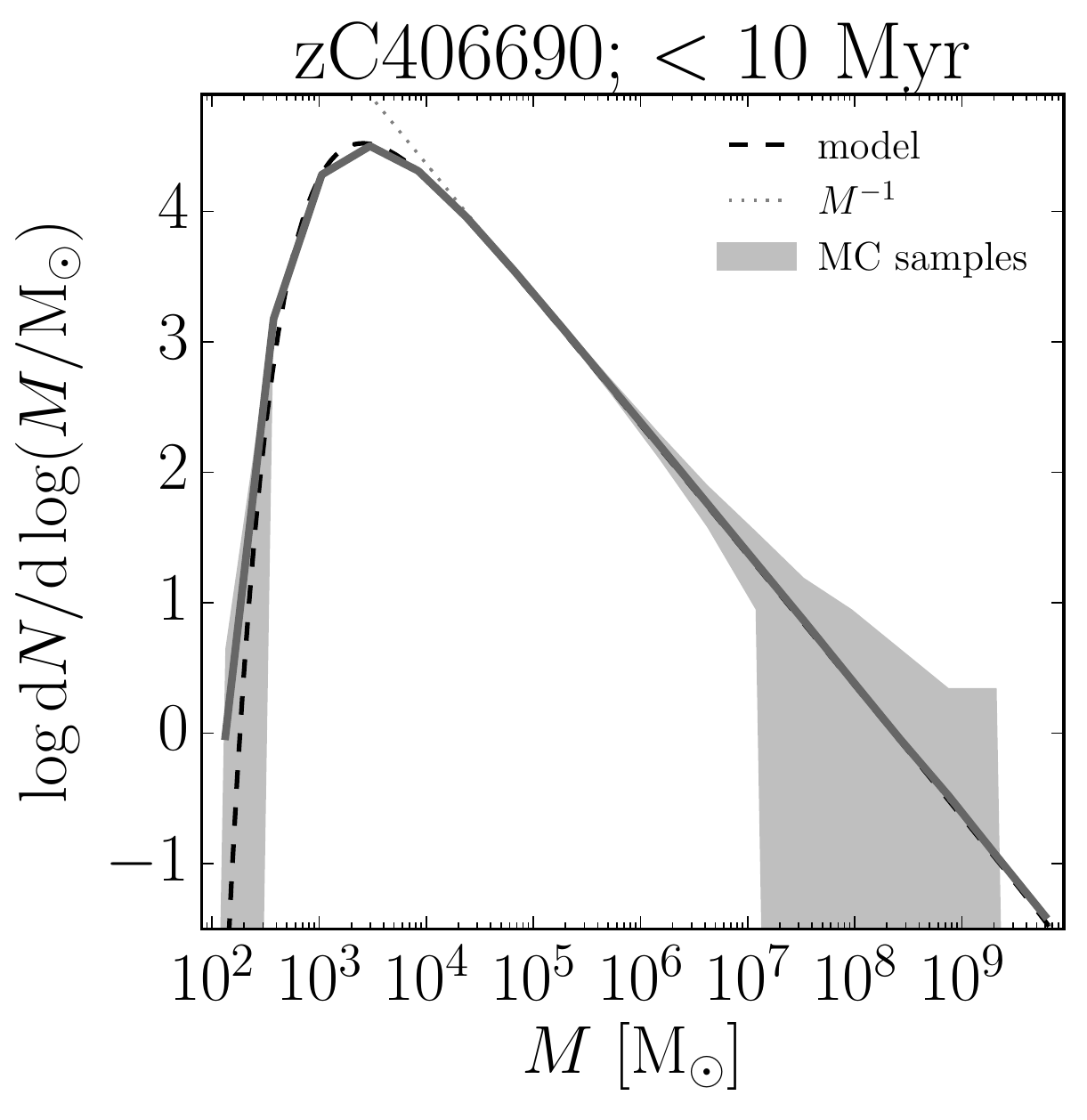}
    \protect\caption{Predicted young CMF in the prototypical high-redshift galaxy zC406690. The prediction of our ICMF model (equation~\ref{eq:CMF}) is shown as a dashed line for clusters with ages $\tau < 5\Myr$. The solid line shows the result of Monte Carlo sampling the predicted ICMF. Because of its high gas surface density ISM, the minimum cluster mass ($\Mmin = 2.6\times10^3\Msun$) is predicted to be more than an order of magnitude larger than in the solar neighbourhood. }
\label{fig:CMF_HZ}
\end{figure}

To obtain the ICMF, we proceed as in Section~\ref{sec:predictions}. We use the rotational velocity and the half-light radius listed in \citet{Tacconi13} (i.e.~$V_{\rm rot} = 224\kms$ and $R_{1/2} = 6.3\kpc$), as well as the peak molecular gas surface density (assuming a molecular gas mass $M_{\rm mol}=8.2\times10^{10}\Msun$ and an exponential gas disc with the same scale-length as the optical disc), $\SigmaISM = 9.3\times10^2\Msunpc2$. In addition, the rotation curve is assumed to be flat at the half-light radius in order to obtain the angular velocity. To calculate the Toomre $Q$ parameter we assume a velocity dispersion of $50\kms$ \citep{Reina-Campos17}. We obtain minimum and maximum masses of $\Mmin = 2.6\times10^3\Msun$ and $\Mmax = 8.9\times10^{10}\Msun$. Figure~\ref{fig:CMF_HZ} shows the predicted ICMF for clusters with ages $\tau < 5\Myr$, as well as a mock measurement and uncertainties obtained with $10^4$ Monte Carlo samples. To normalise the ICMF, we calculate the total mass in clusters using the \citet{Kruijssen12b} model for the CFE and the observed star formation rate, $\rm SFR = 480\Msun~{\rm yr}^{-1}$ \protect\citep{Tacconi13}. The predicted CFE is $\Gamma=68$~per cent. 

To estimate the GCMF that will result from the evolution of the ICMF in zC406690 until the present day, the effects of cluster disruption and dynamical friction must be considered. For instance, \protect\citet{Kruijssen15b} used an analytical treatment to estimate that for a $z=3$ galaxy with $\log(M_*/\Msun) = 10.7$, tidal shocking significantly reduces the number of GCs below $\sim 10^5\Msun$. At the high-mass end, clusters more massive than $\sim 10^6\Msun$ will be depleted by dynamical friction. These combined effects should produce a very narrow GCMF with a peak mass in the range $10^5 - 10^6\Msun$. However, because our model predicts a minimum mass $\Mmin=2.6\times10^3\Msun$, the contribution from disrupted low-mass clusters to the field star population will be reduced somewhat compared to the result of assuming the traditional environmentally-independent $10^2\Msun$ truncation in equation~(\ref{eq:CMF}).

Typical galaxies at $z>1$ have clumpy morphologies in rest-frame UV images. Recent studies use multi-band \emph{HST} photometry to determine the clump mass distributions in  highly-magnified lensed galaxies at $z=1-3$, as well as in deep fields \citep{Adamo13,Elmegreen13,Wuyts14,Dessauges-Zavadsky17,Vanzella17b,Vanzella17a,JohnsonT17}. These studies suggest that the mass function of cluster complexes is truncated above a few times $10^8\Msun$. They also provide an upper limit on the minimum mass of cluster complexes of $\sim 10^{5.5}\Msun$. Considering that, because of the limited resolution, complexes will have masses at least as large or larger than bound clusters, these limits are consistent with our prediction for the extent of the ICMF shown in Figure~\ref{fig:CMF_HZ}. Future studies of lensed high-redshift galaxies with the James Webb Space Telescope and with the next generation of 30-m class ground-based telescopes will probe the CMF in these objects even deeper, allowing for further constraints on GC formation at high redshift.

\section{Conclusions}
\label{sec:conclusions} 

In this paper, we present a model for the environmental dependence of the  minimum mass of bound stellar clusters. The model evaluates the star formation efficiency within feedback-regulated molecular clouds in the context of a rotating galactic disc in hydrostatic equilibrium. In combination with the model for the maximum cluster mass from \citet{Reina-Campos17}, this enables us to predict the full ICMF as a function of the global properties of the host galaxy, namely the surface density of the ISM, the angular velocity of the disc, and the Toomre $Q$ stability parameter. 

We explore the environmental dependence of the minimum cluster mass and the full ICMF in a broad range of galactic environments from local spirals to high-redshift star-forming galaxies, and use it to make predictions for observed young CMFs. The model further allows the reconstruction of the star-forming conditions in local galaxies from their GC populations. Our conclusions are as follows.

\begin{enumerate} 
\item The minimum cluster mass and the resulting total width of the ICMF are predicted to vary by orders of magnitude across the observed range of galaxy properties, from local quiescent discs to high-redshift clumpy star-forming galaxies (Figure~\ref{fig:paramspace}). The main driver of the minimum mass variation is the ISM surface density, with $\Mmin \propto \SigmaISM^3$ for $\SigmaISM \ga 3\times10^2\Msunpc2$, and no dependence on the disc angular speed across most of the parameter space. At very large gas surface densities ($\SigmaISM > 4\times10^3\Msunpc2$), the minimum mass saturates at the maximum mass predicted by \citetalias{Reina-Campos17}, leading to very narrow ICMFs. These overall trends are largely insensitive to the value of Toomre $Q$. 

\item The minimum cluster mass in high ISM surface density environments scales steeply with the assumed value of the star formation efficiency per free-fall time (Section~\ref{sec:uncertainties}). This implies that future observational evidence of a variation in $\Mmin$ will be a sensitive probe of this parameter, which plays a fundamental role in star formation theories.

\item We predict the full ICMF in several environments across parameter space where observational determinations of the young CMF have been performed. Despite large systematic uncertainties in the observations at low cluster masses, the model shows good agreement where the data is most robust, which is generally above $M \sim 10^3\Msun$. Although the observed turnover of the CMF in the solar neighbourhood at low masses is likely due to incompleteness of the cluster catalogues, it matches well the predicted value of the minimum cluster mass of $\Mmin \sim 1.1\times10^2\Msun$. \emph{Gaia} data is expected to considerably reduce the uncertainties. The ICMF in the LMC and in M31 agree well in the power-law regime and high-mass truncation, but the predicted minimum mass lies below the completeness limits of the observations. 

\item The ICMF model predicts considerably larger minimum cluster masses, $\Mmin \ga 10^{2.5}\Msun$, in starbursting environments (due to their ISM surface densities in excess of $10^2\Msunpc2$; see Figure~\ref{fig:paramspace}), and extremely narrow mass functions in high-shear environments found in galactic nuclei. This makes these systems ideal for testing our model. The predicted ICMFs agree very well with the limits set by observations of the Antennae galaxies, as well as in the nucleus of M82. In both cases the model predicts a minimum mass that is several times larger than in the solar neighbourhood, but still below the completeness limits. For the CMZ, we predict the most extreme deviation from the traditional mass function: a narrow ($\sim 1$ dex in mass) peak with a minimum cluster mass of $\Mmin \sim 3.2\times10^3\Msun$. This agrees with the limits set by the masses of young clusters and embedded proto-clusters in the region.

\item The model allows us to predict how the star-forming environments at high redshift shaped the observed GC populations around local galaxies. Conversely, it can be inverted to constrain the star-forming environment of the progenitors of local galaxies during the formation epoch of their GCs. Using this approach, we investigate the possibility that the large GC specific frequency (at low metallicities) in the Fornax dSph could be due to environmental conditions that led to a narrow ICMF at the time its GCs formed. We infer a narrow range of conditions for the ISM surface density and shear in the ISM of the progenitor of Fornax $\sim 10-12\Gyr$ ago. The model predicts that the galaxy must have been quite compact, with ISM surface densities $\SigmaISM > 700 \Msunpc2$ and angular velocities $\Omega > 0.4\Myr^{-1}$ (assuming $Q=0.5$). This implies that the central region was heavily dominated by dense gas and that its stellar component likely underwent considerable expansion to become the spatially extended galaxy that is observed at present.

\item The ICMF models predict that $\sim 80$ per cent of the low-metallicity stars (i.e.~$[{\rm Fe/H}]<-2$) in the Fornax dSph formed in bound clusters, with a large minimum cluster mass of $\Mmin \sim 1.2{-}5.8\times10^3\Msun$. This is more than an order of magnitude larger than the traditionally assumed low-mass ICMF truncation at $10^2\Msun$. The dearth of low-mass clusters at formation explains the puzzling high specific frequency of GCs (relative to metal-poor field stars) observed in dwarf galaxies like Fornax, IKN and WLM by \citet{Larsen12,Larsen14,Larsen18}.
\end{enumerate}

As shown in this paper, modelling the environmental dependence of the CMF has many potential applications for understanding the formation of stellar clusters, as well as for reconstructing the star-forming conditions during galaxy evolution. Future observations of the low-mass regime of the CMF in nearby galaxies with upcoming observational facilities (e.g.~the {\it James Webb Space Telescope} and 30-m class ground-based telescopes) will allow for the model presented here to be tested conclusively. Finally, the model is ideally-suited for implementation in sub-grid models for cluster formation and evolution in (cosmological) simulations of galaxy formation and evolution \citep[e.g.][]{Pfeffer18,Kruijssen19a,li18}. For the foreseeable future, these models will be incapable of resolving the complete stellar cluster population down to the minimum mass scale. The presented model provides a physically-motivated way to account for the low-mass cluster population.

\section*{Acknowledgements}

The authors would like to thank the anonymous referee for a prompt and constructive review. We also thank Angela Adamo, Nate Bastian, Bruce Elmegreen, Cliff Johnson, and Anil Seth for illuminating discussions, as well as Henny Lamers for providing his mass and age estimates of clusters in the solar neighborhood. We gratefully acknowledge funding from the European Research Council (ERC) under the European Union’s Horizon 2020 research and innovation programme via the ERC Starting Grant MUSTANG (grant agreement number 714907). MRC is supported by a Fellowship from the International Max Planck Research School for Astronomy and Cosmic Physics at the University of Heidelberg (IMPRS-HD). JMDK gratefully acknowledges funding from the German Research Foundation (DFG) in the form of an Emmy Noether Research Group (grant number KR4801/1-1).

\bibliographystyle{mnras}
\bibliography{merged}

\appendix

\section{The effect of IMF sampling on the maximum cluster mass}
\label{sec:appendix}

Here we show the effect of IMF sampling in low-mass clouds on the maximum cluster mass model from \citetalias{Reina-Campos17}. Using equations (\ref{eq:tsn_IMF}) and (\ref{eq:delta_t}) to rewrite the SN timescale in equation 4 of \citetalias{Reina-Campos17},  we obtain
\begin{equation}
    t_{\rm fb} = \frac{\tsn}{2} \left[ 1 + \sqrtsign{ 1 + \frac{ 4\pi^2 G^2 t_{\rm ff,g} Q^2 \SigmaISM^2 }{ \phifb \epsff \tsn^2 \kappa^2 }  }  \right] ,
    \label{eq:tfb_Mmax}
\end{equation}
with
\begin{equation}
    \tsn = t_{\rm OB,0} + \frac{ \MOB }{ \epsff } \frac{ t_{\rm ff,g} }{ M_{\rm GMC,max} } ,
    \label{eq:tsn_Mmax}
\end{equation}
where $M_{\rm GMC,max}$ is the maximum mass of a GMC, and $t_{\rm ff,g}$ is the vertical free-fall time of the gas at the midplane. Equation~9 in \citetalias{Reina-Campos17} can then be solved implicitly for the maximum GMC mass that can condense out of the ISM before it is dispersed by SN feedback, delayed by IMF sampling. This yields
\begin{equation}
    M_{\rm GMC,max} = \frac{4\pi^5 G^2 \SigmaISM^3}{\kappa^4} \times \min\left[1, \frac{ t_{\rm fb}\left(M_{\rm GMC,max}\right) }{ t_{\rm ff,2D} }\right]^4 ,
    \label{eq:Mgmcmax}
\end{equation}
where
\begin{equation}
    t_{\rm ff, 2D} = \sqrt{\frac{2\pi}{\kappa}} 
\end{equation}
is the two-dimensional free-fall time of the shear-enclosed sheet of the ISM.

\begin{figure*}
    \includegraphics[width=0.68\textwidth]{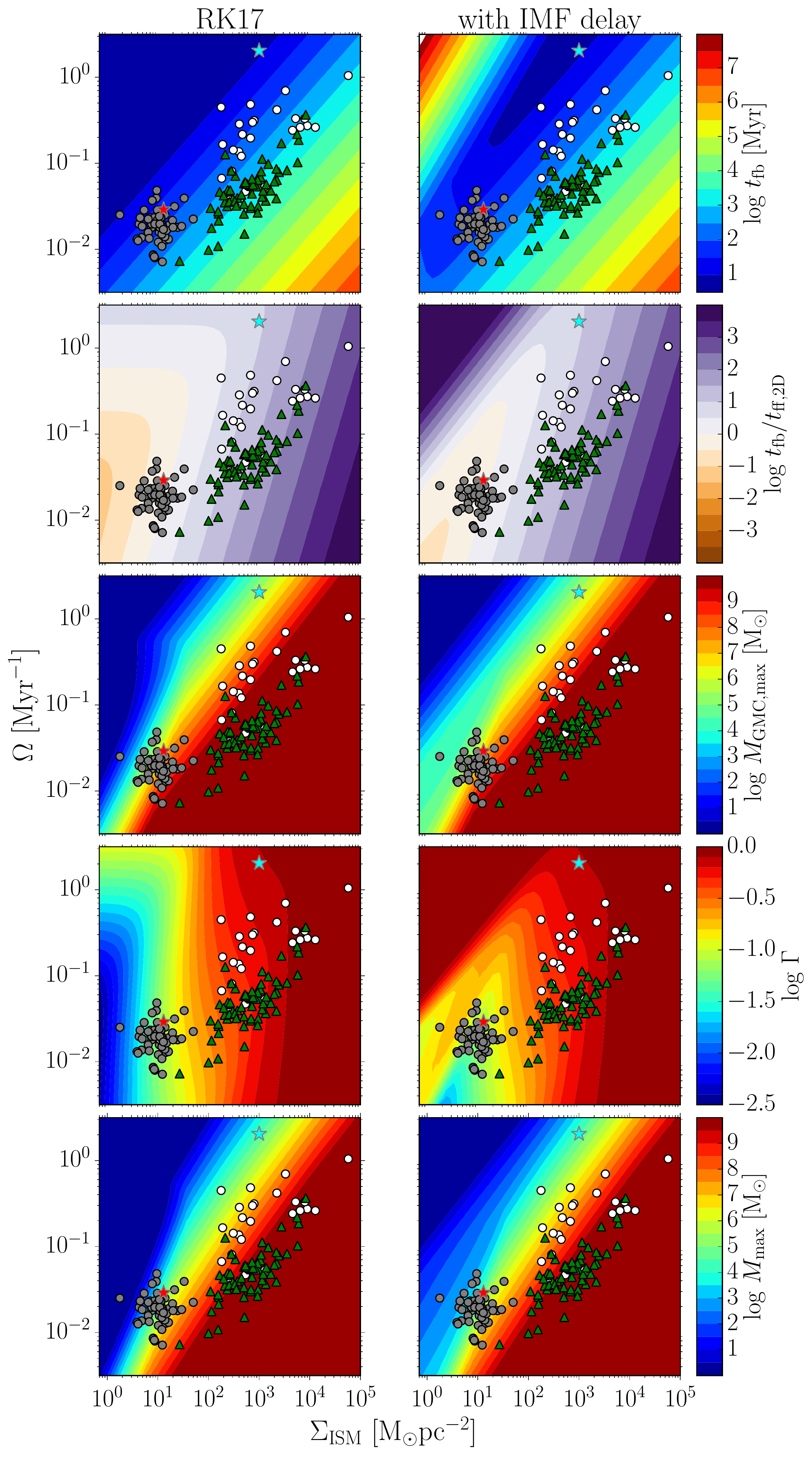}
    \protect\caption{Illustration of the effect of IMF sampling on each element of the maximum mass model assuming $Q=3$. The first column shows the original \citetalias{Reina-Campos17} model, while the second column shows the result of accounting for the feedback delay due to IMF sampling. From top to bottom, the rows show the feedback timescale, the ratio of feedback to free-fall timescale, the maximum GMC mass, the bound fraction of star formation $\Gamma$, and the maximum cluster mass. Within the feedback-limited region (where $t_{\rm fb} < t_{\rm ff,2D}$ IMF sampling imposes a lower limit on the maximum cluster mass $\simeq 100\Msun$ for observed galaxies with the lowest gas surface densities. The colour bar in the second row is limited to the range $t_{\rm fb}/t_{\rm ff,2D}=10^{-4}{-}10^4$, while the colour bars in the third and last rows are limited to the range $M_{\rm GMC,max}=10^0{-}10^{10}\Msun$ for clarity.}
\label{fig:MaxMassIMF}
\end{figure*}
The effect of the delay in the SN detonations due to sampling of the IMF is shown in Figure \ref{fig:MaxMassIMF}, where the feedback timescale, the ratio of the feedback timescale to the 2D free-fall time of the ISM, and the maximum GMC mass are compared with the \citetalias{Reina-Campos17} results. Including the IMF sampling delay has the effect of increasing the feedback timescale at very low ISM surface densities and high angular speeds, where the maximum GMC masses predicted by \citetalias{Reina-Campos17} are small. In this regime, however, the collapse of the clouds is dominated by centrifugal forces. This results in a very small change in the maximum GMC mass with respect to \citetalias{Reina-Campos17}. The GMC masses do increase significantly at very low surface densities $\SigmaISM \la 3\Msunpc2$ and angular speeds $\Omega \la 0.1\Myr^{-1}$. As can be seen in Figure \ref{fig:MaxMassIMF}, most observed galaxies do not occupy this region of the parameter space.

The maximum cluster mass is determined by multiplying the integrated star formation efficiency ($\epsilon$) by the bound fraction of star formation  ($\Gamma$) from \citet{Kruijssen12b} and by the maximum GMC mass, i.e.
\begin{equation}
    \Mmax = \epsilon~\Gamma(\SigmaISM,\kappa,Q)~M_{\rm gmc,max} .
    \label{eq:Mmax}
\end{equation}
Because the \citet{Kruijssen12b} model also relies on the feedback timescale, it must be modified to include IMF sampling in low mass clouds. To do this, we use the feedback timescale of the maximum GMC mass (equation~\ref{eq:tfb_Mmax}) for each position in the parameter space to calculate $\Gamma$. Figure~\ref{fig:MaxMassIMF} shows the effect of this modification on $\Gamma$ and on the maximum cluster mass. Since the bound fraction is weakly time-dependent \citep[see fig.~4 of][]{Kruijssen12b}, we evaluate it at the moment of completion of the star formation process, i.e.\ at $t = t_{\rm fb}$.

While the bound fraction of stars forming in the most massive GMC increases considerably at low surface densities ($\SigmaISM \la 3\Msunpc2$) and at intermediate surface densities and high angular speeds ($\SigmaISM \la 10^2\Msunpc2$ and $\Omega \ga 0.1\Myr^{-1}$), this region is  mostly unoccupied by observed galaxies. The overall effect of the IMF sampling delay on the maximum cluster mass is that of imposing a lower limit on the maximum cluster mass of $\Mmax \geq \MOB \sim 100\Msun$ in the feedback-limited region of the parameter space. This corresponds to the low surface density and angular speed region in the bottom row of Figure~\ref{fig:MaxMassIMF}.

\bsp

\label{lastpage}

\end{document}